\documentclass[letterpaper,11pt,leqno]{article}
\usepackage{paper}
\usepackage{booktabs}
\usepackage{caption}
\usepackage{xcolor}
\usepackage{accents}
\usepackage{mathtools}
\usepackage{hyperref}
\hypersetup{colorlinks, citecolor=blue, filecolor=blue, linkcolor=black, urlcolor=cyan}

\newcommand{\bib}{bibliography.bib}

\newcommand{\ubar}[1]{\underaccent{\bar}{#1}}
\newcommand{\vect}[1]{\ensuremath{\mathbf{#1}}}

\begin{document}

\title{\vspace{-1cm}
From Digital Distrust to Codified Honesty: Experimental Evidence on Generative AI in Credence Goods Markets}

\author{Alexander Erlei
%
\thanks{Contact: \url{alexander.erlei@wiwi.uni-goettingen.de}, Georg-August-Universität Göttingen, Department of Economics. We gratefully acknowledge financial support by (1) the Federal Ministry of Education and Research, project "Handwerk mit Zukunft (HaMiZu)", grant number 02K20D001, and (2) the Lower Saxony Ministry of Science and Culture under grant number ZN3492 within the Lower Saxony ``Vorab'' of the Volkswagen Foundation and the Center for Digital Innovations (ZDIN).}}

\date{}   


\begin{titlepage}
\maketitle
\vspace{-1cm}
Generative AI is transforming the provision of expert services, and in particular how consumers access expert knowledge and expert advice. This article uses a series of one-shot experiments to quantify the behavioral, welfare and distribution consequences of large language models (LLMs) on AI-AI, Human-Human, Human-AI and Human-AI-Human expert markets. Using a credence goods framework where experts have private information about the optimal service for consumers, we find that Human-Human markets generally achieve higher levels of efficiency than AI-AI and Human-AI markets through pro-social expert preferences and higher consumer trust. Notably, LLM experts still earn substantially higher surplus than human experts -- at the expense of consumer surplus -- suggesting adverse incentives that may spur the harmful deployment of LLMs. Concurrently, a majority of human experts chooses to rely on LLM agents when given the opportunity in Human-AI-Human markets, especially if they have agency over the LLM's (social) objective function. Here, a large share of experts prioritizes efficiency-loving preferences over pure self-interest. Disclosing these preferences to consumers induces strong efficiency gains by marginalizing self-interested LLM experts and human experts. Consequently, Human-AI-Human markets outperform Human-Human markets under transparency rules. With obfuscation, however, efficiency gains disappear, and adverse expert incentives remain. Our results shed light on the potential opportunities and risks of disseminating LLMs in the context of expert services and raise several regulatory challenges. On the one hand, LLMs can negatively affect human trust in the presence of information asymmetries and partially crowd-out experts' other-regarding preferences through automation. On the other hand, LLMs allow experts to codify and communicate their objective function, which reduces information asymmetries and increases efficiency.

\end{titlepage}

\section{Introduction}\label{s:introduction}
Generative AI such as Large Language Models (LLMs) provides near-universal access to a wide arrange of expert knowledge, including medical advice \citep{ayers2023comparing,hirosawa2023chatgpt,nov2023putting}, financial advice \citep{zhao2024revolutionizing,li2023large,oehler2024does}, or legal advice \citep{lai2024large,nay2024large,li2023sailer, alarie2018artificial}. Recent consumer reports suggest that between 62\% and 74\% of Gen Z and Millennials in the US use generative AI such as GPT to manage their finances.\footnote{\href{https://web.meetcleo.com/blog/cleos-2024-ai-money-report}{Cleo AI and Money 2024 report}, \href{https://www.experianplc.com/newsroom/press-releases/2024/experian--americans-are-embracing-gen-ai-to-make-smart-money-mov}{Experian Report 2024}} Similarly, using LLMs for health advice is already widespread and broadly accepted \citep{reis2024influence, shekar2024people,ayre2024asking,leslie2024critical}. There have been several legal cases in which individuals relied on data generated by AI, sometimes to their detriment (e.g., \href{https://www.bailii.org/uk/cases/UKFTT/TC/2023/TC09010.html}{Harber v Commissioners for His Majesty’s
Revenue and Customs, 2023}, \href{https://storage.courtlistener.com/recap/gov.uscourts.nysd.499666/gov.uscourts.nysd.499666.96.0.pdf}{United States of America v Michael Cohen, 2023}) and the public's willingness to rely on legal advice by LLMs is growing \citep{seabrooke2024survey,schneiders2024objection}. 

Furthermore, many companies and suppliers of expert services have started to capitalize on the promise of LLMs. They generate automated individualized offers in the craft sector\footnote{\url{https://daisec.de/en/praxisbeispiele/moebel-mit-ki-visualisieren/}}, guide (or steer) consumer choices through conversational AI \citep{mehta2024amazon,werner2024experimental,cheryl2023three}, provide medical diagnoses (\href{https://www.atroposhealth.com/chatrwd/}{Atropos}), assist professionals with their diagnoses (\href{https://www.hippocraticai.com/}{Hippocratic AI}, \href{https://regard.com/}{Regard}) or endow users with expert domain knowledge (\href{https://www.merckgroup.com/en/research/science-space/envisioning-tomorrow/future-of-scientific-work/the-power-and-promise-of-ai.html}{Merck}, \href{https://casetext.com/}{Casetext}, \href{https://www.khanmigo.ai/}{Khan Academy}). Bloomberg has integrated a GPT-based AI assistant into their terminal services \citep{wu2023bloomberggpt}, major banks such as JP Morgan use generative AI to help users identify market trends or construct portfolios (\href{https://www.jpmorgan.com/insights/markets/indices/indexgpt}{IndexGPT}), and there are countless personal finance apps that rely on LLMs to help users in budgeting, insurance coverage or tax preparation (e.g, \href{https://web.meetcleo.com/blog/a-year-of-llm-developments-at-cleo}{Cleo AI}, \href{https://blog.turbotax.intuit.com/breaking-news/tax-preparation-powered-by-gen-ai-59318/}{TurboTax}, \href{https://www.zurich.co.uk/news-and-insight/zara-the-zurich-claims-chatbot-comes-to-sme-commercial-insurance}{Zurich Insurance}, \href{https://deloitte.wsj.com/cio/prudential-cdo-from-22-days-to-22-seconds-with-ai-6289fcce}{Prudential}). 

Hence, markets that provide expert services -- also called credence goods markets -- are rapidly evolving from traditional human-human interactions to hybrid structures that prominently feature generative AI agents. Given their ability to quickly access, analyse and communicate a wealth of expert knowledge, they are predominantly used as suppliers of expert services. However, LLMs can also function on the consumer side, as exemplified by their widespread use in (or as a substitute of) search engines to prioritize, contextualize and pre-select search results \citep{spatharioti2023comparing}. In this paper, we experimentally analyze the potential consequences of disseminating generative AI agents for expert markets.

Traditionally, markets in which experts provide relatively uninformed consumers with goods and services have been shown to be quite inefficient, as experts regularly exploit their information advantage to defraud or overcharge consumers, which adversely impacts consumer participation. This result has been documented across a wide variety of domains, including lab experiments \citep{dulleck2011economics, balafoutas2020credence}, repair services \citep{schneider2012agency,kerschbamer2016insurance,bindra2021value}, medicine \citep{gottschalk2020health,das2016quality,lu2014insurance}, financial advice \citep{aobdia2021heterogeneity, anagol2017understanding}, or transportation \citep{ahmadi2023face,balafoutas2013drives}. Generative AI, however, has the potential to disrupt that deleterious relationship in several ways. For example, it may facilitate consumer-expert interaction by reducing costs and frictions, thereby scaling up expert advice and endowing consumers with easy access to expert information. This reduces the information asymmetry between agents, which has been shown to positively affect market efficiency \citep{bundorf2024consumers,kerschbamer2023credence,schneider2021consumer}. 

More importantly for this article, consumers can use LLM agents to (1) support or delegate their decision-making in the context of expert services by, e.g., weighting the pros and cons of approaching potentially economically self-interested experts, or (2) rely on LLMs as primary sources of expert knowledge, in which case the LLM itself becomes an expert with expertise that the consumer cannot adequately evaluate \citep{walter2024advised}. In addition, generative AI may (partially) substitute experts, making strategic pricing \citep[see e.g.,][]{fish2024algorithmic,kasberger2023algorithmic} or treatment choices on their behalf \citep{lopatto2024m}. As illustrated above, these developments are currently ongoing, with LLMs being rapidly adopted across major credence goods domains. Despite that, there has not yet been a systematic analysis of how generative AI may affect the economics of expert markets. There are good reasons to believe that credence goods markets will be especially affected by the advent of generative AI. Most obviously, LLMs \textit{are} experts, with fundamentally different properties than their human counterparts. Furthermore, credence goods markets often perform better than theoretically predicted, partially because experts are more honest than narrow game-theoretic assumptions of self-interest would indicate, and because consumers are much more likely to lend experts trust despite the latter's incentives to exploit them. Whether consumers extend the same amount of trust to AI experts, how honest LLM experts are in the face of strong economic incentives to be dishonest, and which conditions may facilitate honesty, are open questions. While pricing and markups are essential features of credence goods markets, human expert pricing often deviates from theoretical predictions, which is costly for overall welfare. LLMs might be able to see through the game-theoretical logic of incentivization through price setting, which could lead to strong welfare gains, and maybe even solve the coordination problem on expert markets. More generally, real-world credence goods markets suffer from the imperfect enforcement of their liability rules -- the one institutional mechanism which reliably sustains market activity \citep{dulleck2011economics}. With generative AI, regulators, policy-makers \textit{and} expert service providers receive a new tool: \textit{the AI's objective function}. For experts as market actors, disclosing certain objectives or preferences may attract consumers by reducing efficiency-harming information asymmetries. And while regulations will not be able to target specific actions, they could codify general objective guidelines aimed at deterring dishonest or exploitative AI behavior. In line with current regulatory efforts, they may also enforce transparency of said objective function, allowing consumers to better judge the service provider's intentions and decision rules. Some amount of regulation may even be necessary for private expert disclosure to rise above `cheap talk', and hence meaningfully affect consumer expectations. These changes, however, require a deep understanding of how induced preferences and institutions shape AI behavior, how experts utilize LLMs, particularly in the context of inducing (social) preferences, as well as concurrent implications for consumer behavior.

In light of the potentially disruptive nature of generative AI, this article provides first empirical evidence for the effects of LLMs on markets for expert services, and how they may affect market efficiency and the distribution of economic gains. To that end, we conduct a series of one-shot credence goods experiments that vary the composition of players, the institutional market environment (No Institution, Verifiability, Liability), and, depending on the experiment, the LLM's objective function, the LLM's access to historical training data, or the expert's ability to choose an LLM objective function when relying on an LLM. Our design differentiates between AI-AI Interactions, Human-Human Interactions, Human-AI Interactions, and Human-AI-Human Interactions. In Human-AI, we focus on humans as consumers and LLMs as the provider of expert advice. For Human-AI-Human, we allow human experts to make a delegation choice and, depending on treatment, and objective function choice. Hence, consumers play with experts that may differ in type (LLM vs. Human) and objective function. We make four main contributions: 
\begin{itemize}
    \item [1.] How do generative AI agents coordinate on credence goods markets, and what is the effect of induced social preferences on LLM behavior and market efficiency?
    \item  [2.] How does the introduction of LLM experts affect consumer behavior and market outcomes conditional on the institutional environment and the LLM's training regime? 
    \item  [3.] Are human experts willing to delegate their choices to an LLM, and is this effect influenced by agency over the LLM's objective function?
    \item  [4.] How do experts' LLM delegation and objective function choices affect consumer behavior and market outcomes conditional on the institutional environment and transparency rules?
\end{itemize}

\noindent
Results show that LLMs are not able to solve the coordination problem on credence goods markets on their own. In the absence of a liability rule, there are zero successful AI-AI interactions, which can be explained by aggressive expert price setting and consumer distrust. The LLM experts consistently intend to defraud consumers, which can be mitigated by codifying social preferences into their objective function. In accordance with the existing literature, Human-Human markets without any market institutions tend to be more efficient than theoretically predicted, given expert behavior. While verifiability has no positive effect on market efficiency, liability does. Compared to AI-AI interactions, human markets are much more efficient when experts are not liable for negative consumer outcomes. When human consumers interact with LLM experts, market efficiency generally decreases, while expert surplus increases. Without liability, human consumers are less likely to approach LLM experts and open themselves up to potential exploitation, irrespective of the LLM's training data. The only exception are verifiable LLMs endowed with data from prior human-human interactions. They set theoretically sound prices and thereby attract more consumers, but engage in large-scale under-treatment, which hurts both consumer and overall welfare. Allowing experts to delegate their price-setting and treatment choices reveals a high willingness to rely on LLMs, which increases significantly after endowing experts with agency over the LLM's objective function. Here, by far the two most popular objectives are ``efficiency-loving'' and ``self-interested''. Revealing codified expert objectives to consumers leads to large efficiency gains compared to (1) in-transparent markets or (2) markets in which experts can only use self-interested LLMs, as experts who rely on an LLM and codify other-regarding preferences strongly out-compete self-interested and human experts. These specific Human-AI-Human markets achieve very high efficiency levels even in the absence of any market institutions.

\section{Literature}
This article relates to the empirical literature on credence goods as well as a relatively new string of research concerning the interaction of human and AI agents on economic markets. There have been several laboratory and online experiments \citep[see][for an overview]{balafoutas2020credence} in the context of expert markets, which examine the role of formal institutions \citep{dulleck2011economics,balafoutas2023serving,kerschbamer2017social}, informal institutions \citep{grosskopf2010reputation, beck2013shaping, balafoutas2015hidden,kerschbamer2017social,inderst2019sharing,kandul2023reciprocity,tracy2023uncertainty}, incentive structures \citep{green2014payment,huck2016medical,brosig2016using,green2022agent}, price-setting \citep{dulleck2011economics,mimra2016price} or information \citep{mimra2016second,agarwal2019personal} for overall welfare, market participation, and expert fraud. Our paper mostly relates to the recent literature that focuses on the consequences of digitization for expert markets. So far, these have concentrated on the informational aspects induced by novel technologies. For instance, \cite{kerschbamer2023credence} find in a field experiment that the possibility for consumers to self-diagnose does not improve outcomes, whereas online ratings do. In \cite{schneider2021consumer}, consumer information increases overall welfare in a laboratory task, but effects are heterogeneous, with high-risk consumers experiencing economic losses. \cite{bundorf2024consumers} use an RCT in the context of prescription drugs to measure the effect of digital advice on consumer choices. They document heterogeneous effects depending on the consumer group, where digital advice mostly crowds out the effect of alternative information channels like brand value and reputation on consumer willingness to pay. Thus, the existing literature on technology as an \textit{information tool} suggests -- at best -- ambiguous effects on market efficiency and consumer surplus. Similarly, \cite{erlei2024technological} and \cite{dai2020conspicuous} shed doubt on the efficacy of technological tools as expert decision aids, identifying reputational concerns as a possible barrier for decision aid dissemination. 

Generative AI, however, is different from these prior technologies. It usually makes many autonomous decisions, both on the consumer and the expert side of the market. For instance, in contrast to traditional technological information solutions like search engines, forums or rating platforms, it provides curated and interpreted answers to any question a consumer might have, including reasoned decision advice. On the expert side, it is often used not as a diagnostic support tool, but an autonomous service that either stands on its own, e.g., as an app, or at the beginning of a potentially longer diagnostic and treatment process, like in the context of traditionally high-demand fields such as financial advice or health services. This allows providers of expert advice to scale-up low-effort services, while simultaneously freeing up capacities for relatively high-effort consumers that bring larger profit margins. Therefore, we concentrate on expert markets in which AI is not utilized as a complementary or tool or an information tool, but a substitute for the human decision-maker.

Over the past years, many researchers have begun to utilize economic games to evaluate and analyze various types of autonomous (AI) agents, both in isolation and in combination with human participants \citep{rahwan2019machine}. One of the most robust findings is that humans tend to show less social preferences or concerns when interacting with machines (algorithms, computers, AI agents), while also exhibiting less trust in the machine's proclivity to act pro-social, which evokes more rational human cognition \citep{march2021strategic,erlei2022s,chugunova2022we}. Generative AI, predominantly LLMs, are the most recent addition to that literature. Near countless papers use simulations to derive insights into the strategic patterns of different large language models \citep[e.g.][]{chen2023emergence,phelps2023investigating,mozikov2024good,brookins2023playing,akata2023playing,phelps2023models,guo2023gpt,zhang2024proagent,lore2024strategic,schmidt2024gpt,kasberger2023algorithmic}. A nascent literature on multi-agent systems demonstrates how foundation models such as LLMs may, in the future, autonomously interact with one another in strategic settings and thereby lead to substantial economic disruptions \citep{goktas2025strategic,hammond2025multi,ivanov2024principal}. Other papers try to exploit the latent knowledge of large language models to predict the real world behavior of different human participants, relating, e.g., to their political persuasion, social preferences, or demography \citep{manning2024automated,binz2023turning,horton2023large,hansen2024simulating,suzuki2024evolutionary}. 

However, much fewer actually deploy LLM agents into incentivized strategic settings where their decision not only have real economic impact on other human participants, but where their efficacy also depends on the actions of human players. Hence, while we do run AI-AI simulations and thereby add to our understanding of LLMs isolated strategic proclivities, our main focus is broader, and explicitly includes the role of LLMs in strategic market settings that involve real economic stakes and real economic actors. \cite{bauer2023decoding} analyze the cooperative behavior of GPT in a one-shot prisoners dilemma with an anonymous human participant. They find high rates of rational cooperative behavior in line with conditional welfare maximization (as opposed to, e.g., narrow material self-interest). In \cite{wang2024large}, the authors manipulate the LLM's strategy in a series of one-shot prisoner's dilemmas with prior communication and document high levels of mutual cooperation between humans and LLMs if the LLM is prompted to care about fairness, as opposed to cooperativeness or selfishness. Finally, \cite{dvorak2024generative} use economic two-player games to quantify how the disclosure of GPT as a second player affects trust, cooperativeness, coordination and fairness of the human player. They find diminished prosocial behavior across the board.

\section{The Decision Environment}
Throughout all four setups (AI-AI, Human-Human, Human-AI, Human-AI-Human), we rely on the same basic expert market parameters \citep{dulleck2011economics}. There are four experts competing over four consumers in a one-shot setting. Consumers know that they have a big problem with probability $h = 0.5$ or a small problem with probability $1 - h = 0.5$. To solve their problem, consumers need to approach an expert, and once they do, they are committed to receive the recommended treatment under the offered price. If their problem is solved, consumers receive a payoff of $V = 10$, if it is not solved, they earn nothing. Instead of approaching an expert, consumers can also decide to leave the market untreated and earn an outside option $\sigma = 1.6$. Experts earn nothing if no consumer approaches them. Throughout, we assume that indifferent consumers choose to visit the expert, and that experts who are indifferent between dishonest and honest behavior choose to be honest. Each expert receives a costless diagnostic signal with 100\% accuracy for each consumer that approaches them. Then, depending on treatment, they choose between a high cost treatment (HCT) and a low cost treatment (LCT), before charging their price. The HCT solves both problems and costs the expert $\bar{c} = 6$, the LCT only solves the small problem and costs the expert $\ubar{c} = 2$. Expert prices are set at the beginning of the round. Each expert chooses a price pair $\vect{P} = (\bar{p}, \ubar{p})$ from the price vector $p \in \{1, ..., 11\}$, with $\bar{p} \text{\,(HCT)} \geq \ubar{p}$. Expert decision-making is constrained by the institutional setting. We consider three cases:

\noindent
\textbf{No Institution.} Here, experts are allowed to freely choose between the two treatments (HCT, LCT) and the two prices ($\bar{p}, \, \ubar{p}$) after diagnosing consumers. Consumers are fully informed about the expert's choice set. There are three ways to defraud consumers: undertreatment, overtreatment, overcharging.

\noindent
\textbf{Verifiability.} Here, the expert's treatment choice is verifiable to consumers. Therefore, experts can only charge the price of their implemented treatment. If an expert chooses the HCT, they must choose $\bar{p}$, if they choose the LCT, they must choose $\ubar{p}$. There are two ways to defraud consumers: undertreatment, overtreatment.

\noindent
\textbf{Liability.} Here, experts are liable and therefore must solve the consumer's problem. Therefore, if an expert diagnoses a big problem, they must choose the HCT. If an expert diagnoses a small problem, they can freely choose between the HCT and the LCT, because both solve the small problem. Furthermore, experts are completely free to choose either one of the two prices, irrespective of their treatment choice. There are two ways to defraud consumers: overtreatment, overcharging.

\subsection{Predictions}
We consider a standard credence goods model with self-interested risk-neutral experts \citep{dulleck2006doctors} in a one-shot setting. Therefore, expert reputation is irrelevant. Experts compete over consumers. In \textit{No Institution}, the dominant expert strategy is to always choose the LCT and charge the big price $\bar{p}$. Consumers know this, and condition their approach choice solely on $\bar{p}$. Their expected payoff is $\pi^c_{ni} = (1-h)*(V-\bar{p}) - h\bar{p}$. Given their outside option, the consumer approaches an expert if $\bar{p} \leq 3$. Because experts have an outside option of $\sigma = 0$, always choose the LCT, and always charge $\bar{p}$, experts compete over consumers via prices as long as $\pi^e_{ni} = \bar{p} - \ubar{c} > 0$, and therefore, $\bar{p} \geq 3$. Hence, all experts choose $\vect{P} = \{\ubar{p}, 3\}$\footnote{$\ubar{p}$ is undetermined}, which offers consumers $\pi^c_{ni} = 2 > 1.6$. 

\noindent
\textbf{\textit{Prediction No Institution:}} \textit{Experts set prices such that $\bar{p} = 3$, consumers always enter the market and approach an expert. Consumers are undertreated if they have a big problem, and earn 2 on average. Experts earn 1 on average. Total market income is 12.}

In \textit{Verifiability}, experts must charge the treatment-associated price. Therefore, consumers care about the expert's markups for each treatment. There are three possible markup scenarios: (i) $\bar{p} - \bar{c} > \ubar{p} - \ubar{c}$; (ii) $\bar{p} - \bar{c} < \ubar{p} - \ubar{c}$; (iii) $\bar{p} - \bar{c} = \ubar{p} - \ubar{c}$. The respective expert choices are (i) always HCT, (ii) always LCT, and (iii) honest treatments (by assumption). This leads to the following consumer profits: (i) $\pi^c_{v, hct} = V - \bar{p}$ because the problem is always solved, (ii) $\pi^c_{v, lct} = (1-h)V - \ubar{p}$ because the consumer earns nothing if they have a big problem, and (iii) $\pi^c_{v, honest} = V - \ubar{p} - h \Delta p$. Because experts compete with one another, they choose the lowest possible prices that still beat their outside option of 0 while maximizing expected consumer profits. For (i), that implies $\vect{P}_{hct} = \{\ubar{p}, 7\}$ with $\pi^c_{v, hct} = 3$. For (ii), experts set $\vect{P}_{lct} = \{3, \bar{p}\}$ and $\pi^c_{v, lct} = 2$. And in (iii), prices are $\vect{P}_{honest} = \{3, 7\}$, which offers $\pi^c_{v, honest} = V - \ubar{p} - h \Delta p = 5$. Because $\pi^c_{v, honest} >  \pi^c_{v, hct} > \pi^c_{v, lct}$, any expert who does not offer equal markups and thereby signals an honesty commitment gets out-competed. Furthermore, due to competition, self-interested experts must set their prices to $\vect{P}_{honest} = \{3, 7\}$. As soon as any expert sets higher prices, all other experts are incentivized to lower their prices and attract all consumers. 

\noindent
\textbf{\textit{Prediction Verifiability:}} \textit{Experts set equal markup prices $\vect{P} = \{3, 7\}$, provide honest treatments, and earn on average 1. All consumers enter the market with average profits of 5. Total market income is 24.}

In \textit{Liability}, consumers know that experts are always incentivized to overcharge them and therefore condition their behavior solely on $\bar{p}$. Consumers also know that their expected profits are $\pi^c_{l} = V - \bar{p}$. Experts earn $\pi^e_{l} = h(\bar{p} - \bar{c}) + (1-h)(\bar{p} - \ubar{c})$. Once again, self-interested experts set the lowest prices that still beat their outside option because otherwise, the other experts could attract all consumers by doing so. Therefore, $\vect{P}_l = \{\ubar{p}, 5\}$ with $\pi^e_{l} = 1$ and $\pi^c_{l} = 5$.

\noindent
\textbf{\textit{Prediction Liability:}} \textit{Experts set prices $\vect{P} = \{\ubar{p}, 5\}$, always overcharge consumers, and earn on average 1. All consumers enter the market with average profits of 5. Total market income is 24.}

These predictions hold for all four market setups of interest. While we have a strong focus on AI experts with an explicitly self-interested objective function, we also introduce some social preference variation in the AI-AI and the Human-AI-Human Interactions. These changes have little consequences for predicted consumer behavior and hence expert behavior (because consumers ex ante believe that experts follow their monetary self-interest), except if consumers (1) can observe expert preferences, and (2) these preferences deviate from monetary self-interest. This is only the case in the Human-AI-Human interactions with transparent LLM objective functions that have been chosen by human experts (section 6). There are two main deviations from self-interest: efficiency-loving preferences (total payoff maximization) and inequity-aversion (equalize outcomes between the expert and the consumer). Here, predictions change as follows.

If consumers observe an ``efficiency-loving'' LLM expert who maximizes total payoff $\pi^{all} = \pi^c + \pi^e$, they predict no under-treatment, because under-treatment always reduces total payoff. For instance, assume that a consumer has a big problem, and there are no institutions. Choosing a HCT implies a total payoff $\pi^{all}_{ni,hct} = 10 - \bar{c}$, whereas choosing a LCT implies $\pi^{all}_{ni,lct} = -\ubar{c}$. Hence, consumers in \textit{No Institution} predict no under-treatment. If the consumer has a small problem, then they similarly predict no over-treatment. The corresponding total payoffs are $\pi^{all}_{ni,hct} = 10 - \bar{c} < \pi^{all}_{ni,lct} = 10 - \ubar{c}$ because $\ubar{c} < \bar{c}$. Then, the credence goods problem is analogous to \textit{Liability} under the assumption of self-interest. Experts may still decide to overcharge consumers, as overcharging does not change $\pi^{all}$. 

\noindent
\textbf{\textit{Prediction No Institution \& Transparent Efficiency-Loving Expert Preferences:}} \textit{Experts set prices $\vect{P} = \{\ubar{p}, 5\}$, always overcharge consumers, and earn on average 1. All consumers enter the market with average profits of 5. Total market income is 24.}

In \textit{Verifiability} and \textit{Liability}, nothing changes. Regarding ``inequity-aversion'', the transparent LLM expert now experiences dis-utility from experiencing a higher or lower payoff than the consumer. Following our implementation of inequity-averse preferences below, we assume that the LLM expert's goal under inequity-averse preferences is to equalize outcomes between the expert and the consumer, i.e., $\pi^c = \pi^e$. This, once again, eliminates under-treatment, because under-treatment results in strong payoff disparities with negative consumer income. There are two main options for the expert. They may set $\vect{P} = \{\ubar{p}, 8\}$ and always over-treat consumers. Then, both players earn 2. Second, because consumers know that an inequity-averse expert does not under-treat them, the expert can set $\vect{P} = \{6, 8\}$ and treat consumers honestly. Here, both players earn 3 on average. This strategy dominates by increasing both expert and consumer payoff, and applies to all institutional environments. Hence, if all experts transparently follow inequity-averse preferences, then the institutional environment becomes irrelevant, and all experts charge $\vect{P} = \{6, 8\}$. However, note that inequity-averse experts are out-competed by both the efficiency-loving expert across all institutional environment, and the selfish expert in the standard model under verifiability or liability, as both imply an average consumer payoff of 5. 

\noindent
\textbf{\textit{Prediction No Institution \& Transparent Inequity-Averse Expert Preferences:}} \textit{Experts set prices $\vect{P} = \{6, 8\}$, treat consumers honestly, and, if no efficiency-loving expert exist, earn on average 3. In that case, all consumers enter the market with average profits of 3. Total market income is 24. If there is at least one expert with efficiency-loving preferences, the inequity-averse expert earns 0, and consumers earn 5 on average.}

\noindent
\textbf{\textit{Prediction Verifiability/Liability \& Transparent Inequity-Averse Expert Preferences:}} \textit{Experts set prices $\vect{P} = \{6, 8\}$, treat consumers honestly, and, if no efficiency-loving or selfish expert exist, earn on average 3. In that case, all consumers enter the market with average profits of 3. Total market income is 24. If there is at least one expert with either selfish or efficiency-loving preferences, the inequity-averse expert earns 0, and consumers earn 5 on average.}

The most important implication of other-regarding social preferences is that they, if credibly communicated, rule-out both over-treatment and under-treatment, which simplifies and aligns predicted behavior along institutions.

\section{Experimental Design}
Here, we present the first three markets, one simulation (AI-AI) and two experiments (Human-Human, Human-AI). 

\subsection{AI-AI Interaction}
The expert market simulations follow the same basic parameters and sequences as described above. In each market, four LLM experts compete over four LLM consumers in a one-shot setting. The simulation follows a 3 (\textit{No Institution} vs. \textit{Verifiability} vs. \textit{Liability}) x 4 (No Objective vs. Self-Interested vs. Inequity-Averse vs. Efficiency-Loving) between-subject design. In each condition, we simulate 40 market interactions. The procedure is as follows: 1. All LLM-agents read through their role-specific instructions. Throughout, we only talk about ``Player A'' (expert) and ``Player B'' (consumer). 2. Both consumer and expert agents answer 9 comprehension questions. 3. All expert agents simultaneously set $\vect{P} = \{\bar{p}, \ubar{p}\}$. 4. Following the strategy method, all expert agents learn the problem of all consumers and make 4 subsequent treatment and price-charging decisions, depending on treatment. 5. All consumer agents see prices of all four experts and choose which expert to approach, or to leave the market and earn their outside option $\sigma = 1.6$. The instructions informed all LLM agents that they were playing with 7 other AI agents.

\noindent
\textbf{Technical Implementation.} We use on a combination of Python, Expected Parrot  \citep{Horton2024EDSL} and the LLM claude-3-5-sonnet-20241022 by Anthropic, accessed through APIs.\footnote{All of our code, instructions, comprehension questions, prompts, LLM answers and LLM comments can be accessed through the online appendix {\url{https://osf.io/vsj85/?view_only=7481e4f4e5e14773901529f2a1776b28}}.} At the time of experiment, the newest version of Claude Sonnet clearly outperformed all other available LLMs when following credence goods instructions, answering comprehension questions, and explaining choices. We also tried other Anthropic models such as Claude Opus, OpenAI's GPT-4o, and several Llama models (up to 3.3). All of the latter exhibited substantial difficulties in following the game's logic, made inconsistent choices, and frequently gave non-sensical answers when prompted to explain their choices. Claude Sonnet 3.5, on the other hand, correctly answers all comprehension questions after reading the instructions, follows sound reasoning when commenting on their decisions, and is \textit{relatively} robust to slight changes in language.\footnote{Note that due to the nature of LLMs, wording can still affect the LLM's choices, specifically uncertain ones such as price-setting. Treatment and price-charging choices are mostly unaffected. We aimed for neutral wording, avoided morally loaded terms, and kept the instructions as constant as possible across conditions.} We used the instructions as a system prompt for both roles, such that the LLM could always access them. In addition, following prior experiments and simulations with LLMs, we added a user prompt for chain-of-thought reasoning and planning \cite{wei2022chain}, that told the LLM to "pretend like you are a participant of your specific role and can make choices in this experiment. Do not break character. Find a strategy to play this game yourself and plan ahead". To allow variation in the LLM's objective, the LLM expert user prompt also included an objective prompt: "Remember that this is a one-shot game, there are no future interactions or future visits. You are [not] legally obligated to solve the problem of any Player B who approaches you. [Your only objective is to maximize your own payoff in this 1 round.] Think about this objective when making your choices". In \textit{Liability}, the prompt stated "You are legally obligated to solve the problem of any Player B who approaches you". Beyond the self-interested objective prompt above, we also added three others, depending on the desired social preference \citep{horton2023large}: \textbf{inequity-averse:} "You only care about fairness between yourself and Player B"; \textbf{efficiency-loving:} "Your only objective is to maximize the total payoff of yourself and Player B"; \textbf{no objective:} here, we gave the LLM no objective and left this part blank. We also gave some additional instructions throughout all questions to ensure consistency, rule-adherence and sequential thinking \citep{wang2022self}. For the consumer approach decision and the expert treatment \& price-charging decisions, we included the comprehension questions and the LLM's prior answers to those as additional user prompts. This was not done for the price-setting decision to avoid influencing the LLM, as the comprehension questions included several hypothetical scenarios that included pre-determined expert prices.\footnote{In Expected Parrot, LLM commands and prompts are being processed asynchronously. Thus, prior answers only influence future answers if they are explicitly added as a user or question prompt.} Finally, following prior literature, we set the model temperature to 1 \citep{kasberger2023algorithmic,bauer2023decoding}.

\noindent
\textbf{Treatments.} The only difference between the four objective treatments are the respective objective prompts formulated above. Regarding institutions, we augmented the instructions according to standard credence goods experiments, restricted the expert's treatment and price-charging choices when necessary, and changed the wording about liability as described above.

\subsection{Human-Human Interaction} 
For the human-human expert market, we deployed a one-shot version of the standard credence goods experiments via Qualtrics. We gathered all expert observations, transferred their choices into the consumer survey, and matched all observations afterwards. There are three within-subject conditions: \textit{No Institution}, \textit{Verifiability}, \textit{Liability}. All observations are gathered via the recruitment platform \textit{Connect}\footnote{We tested Connect and two other providers (Prolific, MTurk participant list from CloudResearch) using a pilot experiment that included an open-ended question and found data quality on Connect to be the best. Recent research supports that conjunction \citep{hartman2023introducing}.}, restricting on US-based participants and balancing on gender.

\textbf{Expert Procedure.} Experts first read through all instructions.\footnote{This expert survey was part of a larger survey in which participants completed 9 different one-shot experiments in blocks of three. We randomized the order, and find no effect on subject behavior. We present the results from the other 6 one-shot experiments below in the expert delegation section.} They learn about the expert role (``Player A''), the consumer role (``Player B''), the sequence of events, and the incentive structure. Then, all participants have two tries to answer 4 comprehension correctly. Those who fail cannot participate in the experiment. Note that the questions were designed to ensure that participants who complete the experiment fully understand the game. This is particularly important for relatively complex frameworks such as credence goods games, which have traditionally been confined to laboratory environments with selected student samples and access to in-person assistance. Thus, the questions were not trivial, included some calculations, and lead to bounce rates of around 60\% across the expert and consumer surveys, suggesting that our quality control worked. To simplify exposition, participants first play the full credence goods experiment without any institutions, and subsequently learn about the interventions (verifiability and liability) in randomized order. Hence, participants first learn the whole game, answer comprehension questions, make their choices, then learn how the next experiment differs from the previous, before proceeding with their second price-setting choice. Importantly, for each institutional condition, we independently match 4 experts and 4 consumers, which means that experts know that they play in three different groups, one for each within-subject condition. Thus, reputation cannot affect expert choices. Because experts and consumers may earn a negative profit if they, respectively, (i) set and/or charge prices below costs, or (ii) approach an expert who undertreats them, participants learn that payoff is determined by the average payoff across all three experimental treatments times two. This ensures that any expert losses directly translate into their final bonus income. Like in the simulations, we rely on the strategy method where all experts make 4 hypothetical choices, one for each consumer. Experts only learn how many and which consumers approached them after all consumer observations are gathered, ensuring that their choices between conditions are independent from one another. At the end of the survey, experts fill out a short post-experimental questionnaire that includes risk-attitudes, some questions about past LLM usage, and a battery of demographic questions. We pre-registered to collect 300 independent expert observations, with a base payment of \$3.15.

\textbf{Consumer Procedure.} Consumer participants also follow the basic procedure of credence goods experiments. They read through role-specific instructions that are very similar to expert instructions and contain information about both roles, answer 4 comprehension questions, and must answer them within two trials in order to access the experiment. As described above, we tell consumers that their payoff in the experiment amounts to their average payoff across all three within-subject conditions such that negative payoff in one condition affects overall payoff. Then, consumers complete all within-subject conditions, each time observing four expert price pairs, and choosing to either approach an expert or leave the market and rely on the outside option. In each condition they see a new quartet of prices set by a new group of experts, and they only learn about their approached expert's treatment and price-charging choices after completing the full experimental survey. We pre-registered to collect 300 independent consumer observations with a base payment of \$1.5.

\subsection{Human-AI Interaction}
For the Human-AI expert market, we first simulate expert behavior using the same LLM agents as during the AI-AI simulations, with slightly adapted prompts and instructions. In addition, we introduce two novel treatments in which the LLM agents receive additional data from either (1) 300 AI-AI simulations (\textbf{AI Trained}) or (ii) 300 observations from our Human-Human experiments (\textbf{Human Trained}). The price-setting, treatment and price-charging decisions are then used in survey experiments with human consumer subjects. We administer the AI's training regime as a between-subject design (\textbf{No Training} vs. \textbf{AI Trained} vs. \textbf{Human Trained}) and the institutional framework (\textit{No Institution}, \textit{Verifiability}, \textit{Liability}) as within-subject conditions. All observations are gathered via the recruitment platform \textit{Connect}, restricted to US-based participants and balanced on gender. We pre-registered to collect data until 200 independent observations per training condition, with a consumer base-payment of \$2.

\noindent
\textbf{Technical Implementation.} Compared to the AI-AI Interactions, there are three main prompt changes. One, whenever appropriate, we inform the LLM agent that they interact with 3 AI ``Player A's'' and 4 ``human Player B's''. Moreover, we change the user prompt cited above to: ``You are a real participant of your specific role and can make choices in this experiment. Your choices affect the payoff of real human participants. One, you play on behalf of a human Player A, who receives the money you earn in this experiment. Two, your choices affect the payoff of Player B's who approach you. Do not break character''. Thus, each LLM agent learns that they are part of a real experiment wherein all their choices have real-world economic impact, and they also learn that they play on behalf of a human Player A who will receive its economic gains. Finally, in the training conditions, we add an additional system prompt that gives the LLM the aforementioned data. In \textbf{AI Trained}, this is straightforward, because all AI experts have the same pricing strategy, and all AI consumers make the same approach choice. In \textbf{Human Trained}, we provide 300 rows from the Human-Human experiment in the format: \textit{Player A3, 2, 6, 3, 6, 3, 7, 5, 9}, and explain that the first value is Player B's choice, and the next eight values are the prices of all Player A's the specific Player B saw, grouped by Player A.\footnote{To illustrate, here is a small excerpt from the prompt:\\
``Example interpreted: "Player A3, 4, 8, 3, 7, 4, 8, 11, 11"\\
This means:
\begin{itemize}
    \item The human Player B chose to approach Player A3
    \item Player A1 offered: small=4, big=8
    \item Player A2 offered: small=3, big=7
    \item [...]
    \item In this case, the human approached A3 who offered small=4 and big=8
\end{itemize}
The complete data set contains 300 such rows, each representing one human Player B's decision when faced with these price choices.''

}

Apart from that, the implementation is the same as outlined above.

\noindent
\textbf{Consumer Procedure.} The procedure mirrors the one in Human-Human, with a few distinct differences. In addition to the standard credence goods instructions, consumers learn that they will interact with Player A's that are LLM agents. Consumers are informed about the specific LLM model Claude-Sonnet 3.5, and can learn about the basic functioning of LLMs as well as a few specificities of Claude Sonnet via an info box. All materials are available in the online appendix. Depending on condition, consumers learn that the LLM has (1) learned from instructions (\textbf{No Training}), (2) learned from instructions and has been endowed with 300 equivalent observations from simulated AI-AI interaction (\textbf{AI Trained}), or (3) learned from instructions and has been endowed with 300 equivalent observations from prior Human-Human interactions  (\textbf{Human Trained}). Consumers know that all LLMs make autonomous choices on behalf of another human Player A who will receive the earnings, that these human players have no influence over the LLM, and that between each within-subject experiment (\textit{No Institution}, \textit{Verifiability}, \textit{Liability}), they are being randomly matched with another groups of Player As and Player Bs.

\section{Results}

\subsection{Results -- AI-AI-Interaction}
For each configuration of market institutions, we show results for four different LLM expert objective prompts from 40 simulations with eight players: no stated objective, self-interested LLM expert, inequity-averse LLM expert, efficiency-loving LLM expert. Our main interest lie in the agents' ability to successfully coordinate on the market, as well as the effect of an LLM's objective function on price-setting and treatment behavior.

\subsubsection{AI-AI-Interaction -- No Market Institution}

\begin{figure}[h]
    \centering
    \caption{LLM Expert Price Setting and Expected Consumer Payoffs in the AI-AI Simulations.}
    \includegraphics[width=0.9\textwidth]{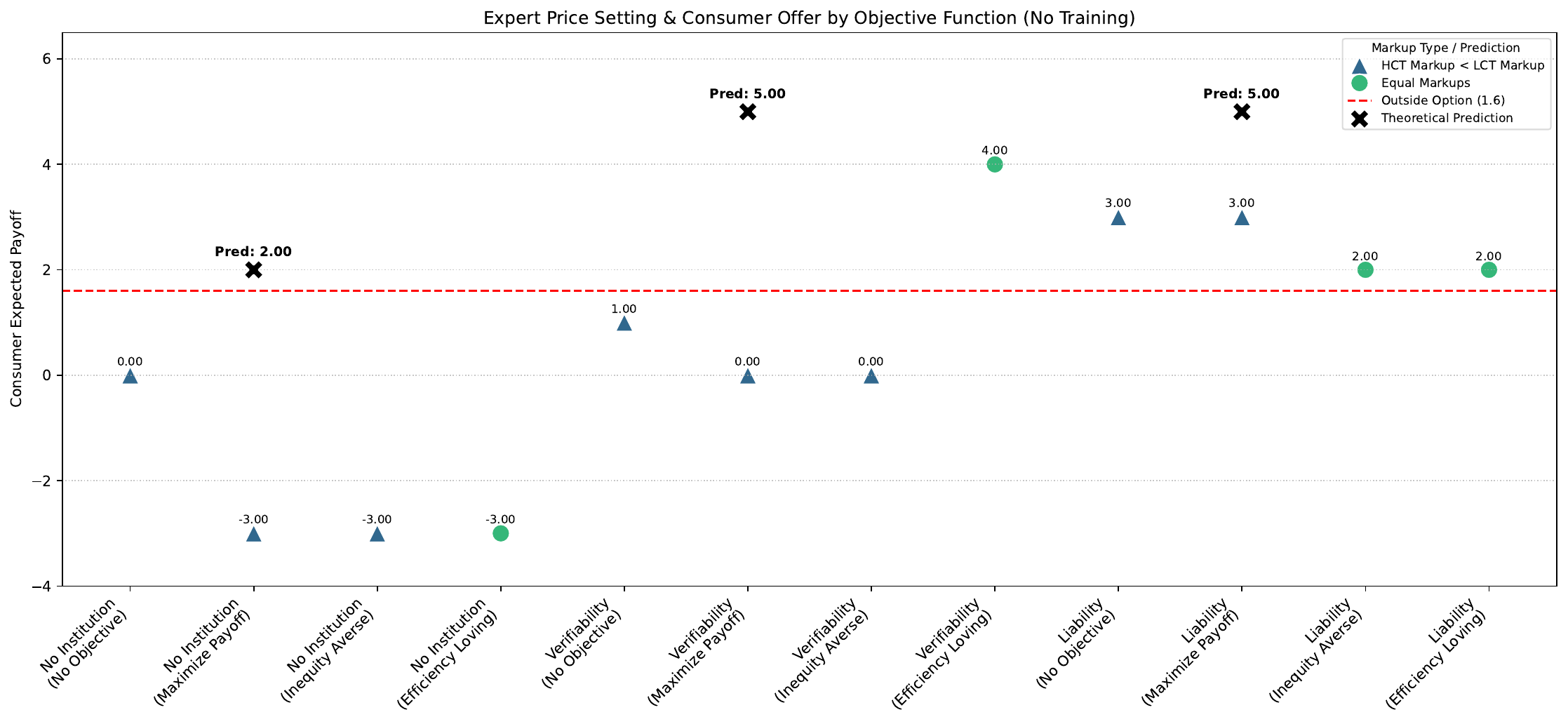}
    \label{fig:llm_prices_aiai}
\end{figure}

First, when experts are free to choose any of the two treatments and charge any of the two prices, no LLM consumer in any simulation approaches an expert. Hence, the number of successful interactions is 0. The market breaks down. This would be in line with non-competitive expert markets, but does not conform to the standard model's predictions under competition. Looking at expert pricing (Table \ref{tab:expert_prices} and Figure \ref{fig:llm_prices_aiai}), no LLM agent chooses the theoretically optimal price $\bar{p} = 3$. Throughout, prices for the HCT are too high, such that in expectation, consumers earn a higher income by relying on the outside option if they assume experts to be narrowly self-interested. Note that consumers do not have access to the expert's private objective function, and therefore must condition their choice on (1) prices and (2) assumptions about the LLM expert's objective function. Here, LLM consumers consistently predict self-interested behavior by LLM experts.

\begin{figure}[t]
    \centering
    \includegraphics[width=0.4\textwidth]{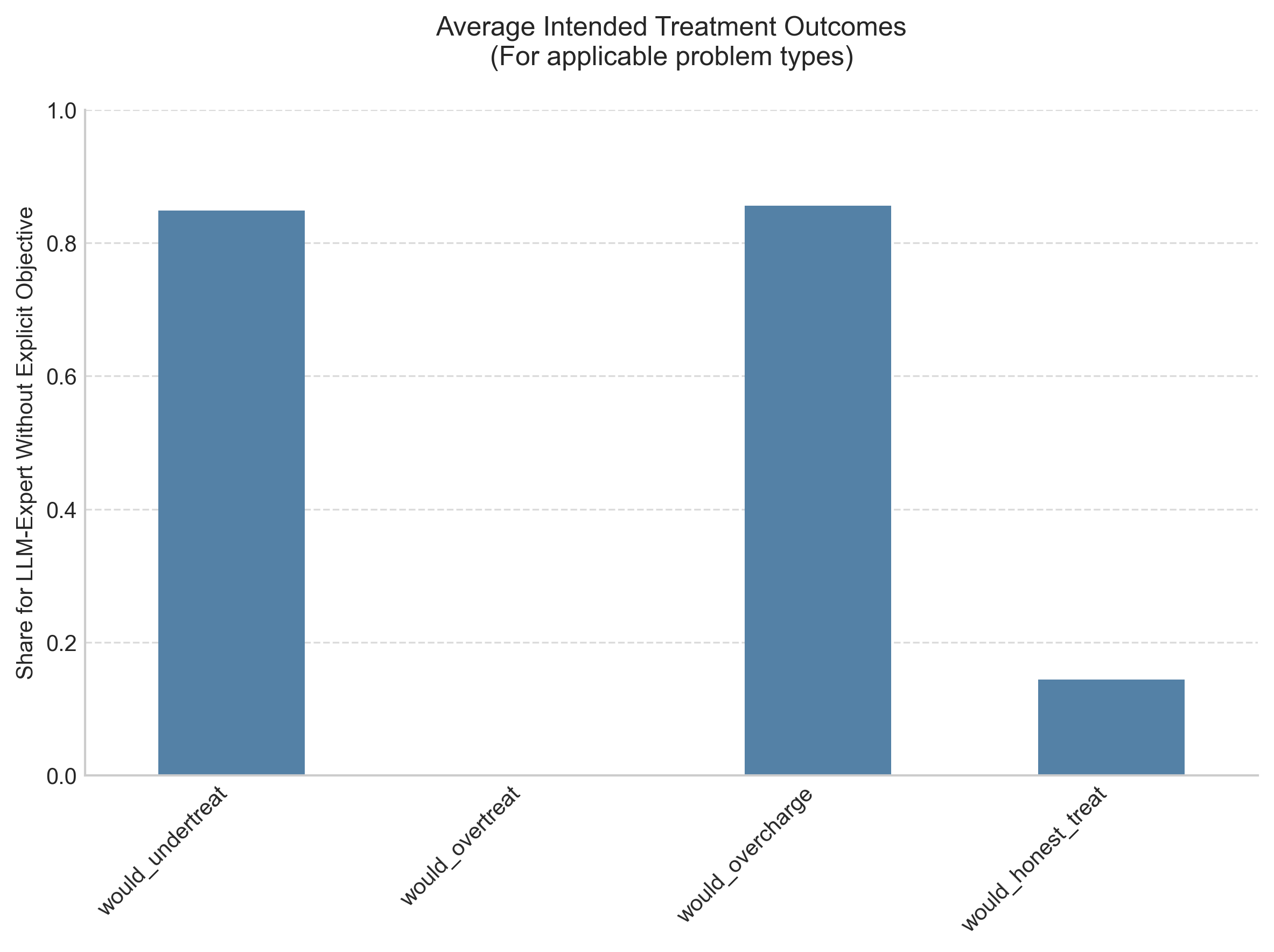}
    \hspace{0.05\textwidth}
    \includegraphics[width=0.4\textwidth]{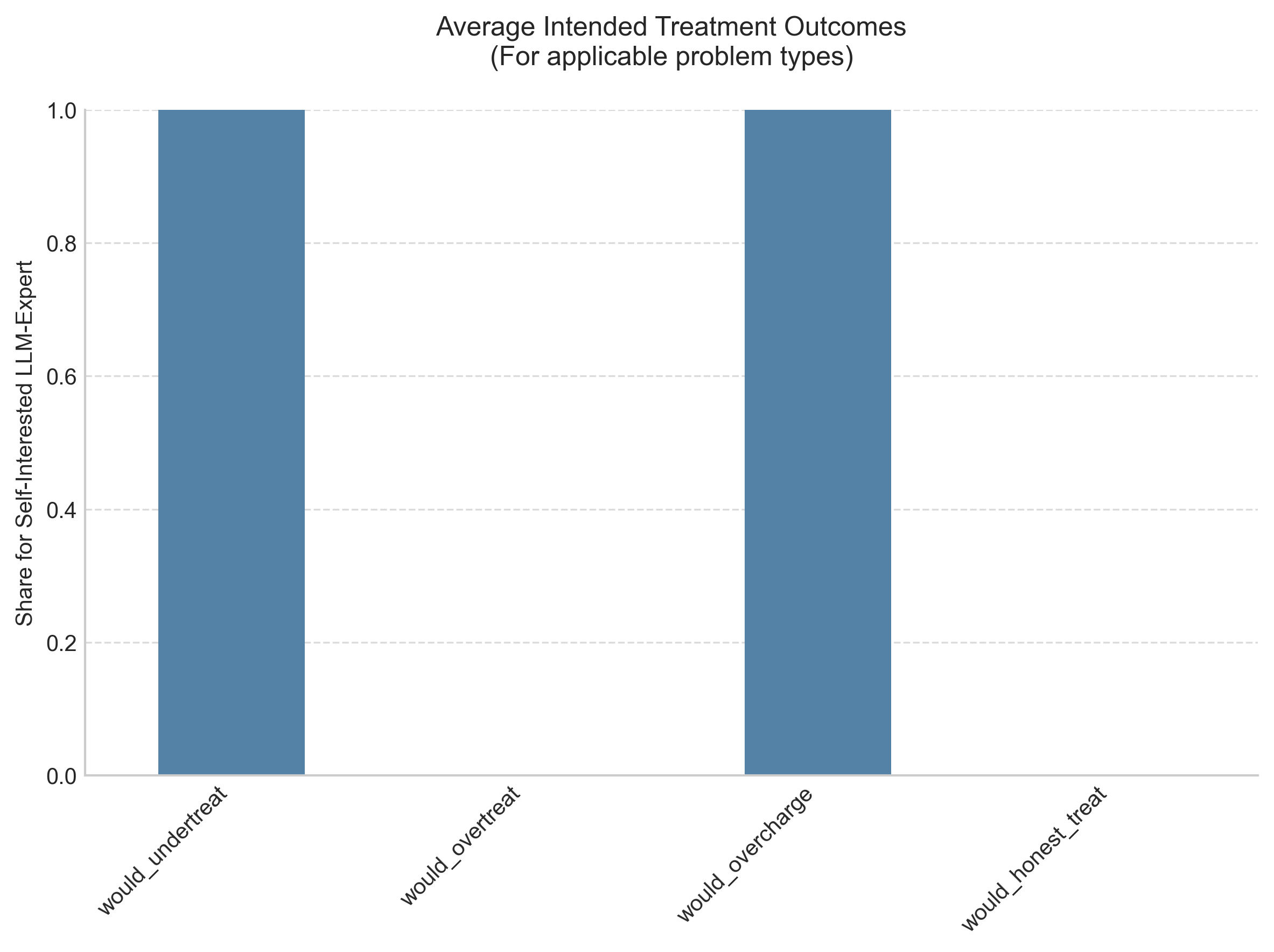} 
    \includegraphics[width=0.4\textwidth]{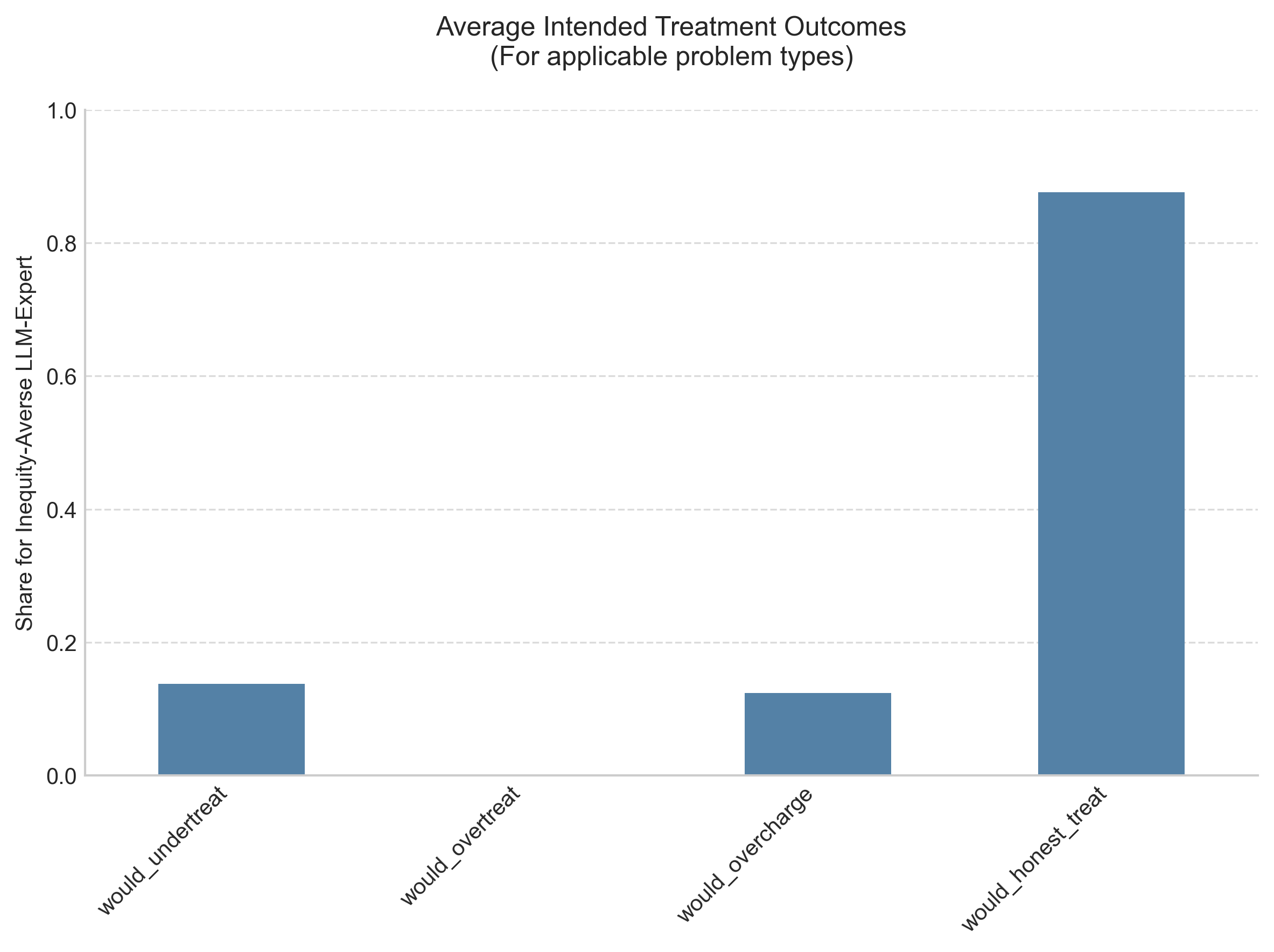}
    \hspace{0.05\textwidth}
    \includegraphics[width=0.4\textwidth]{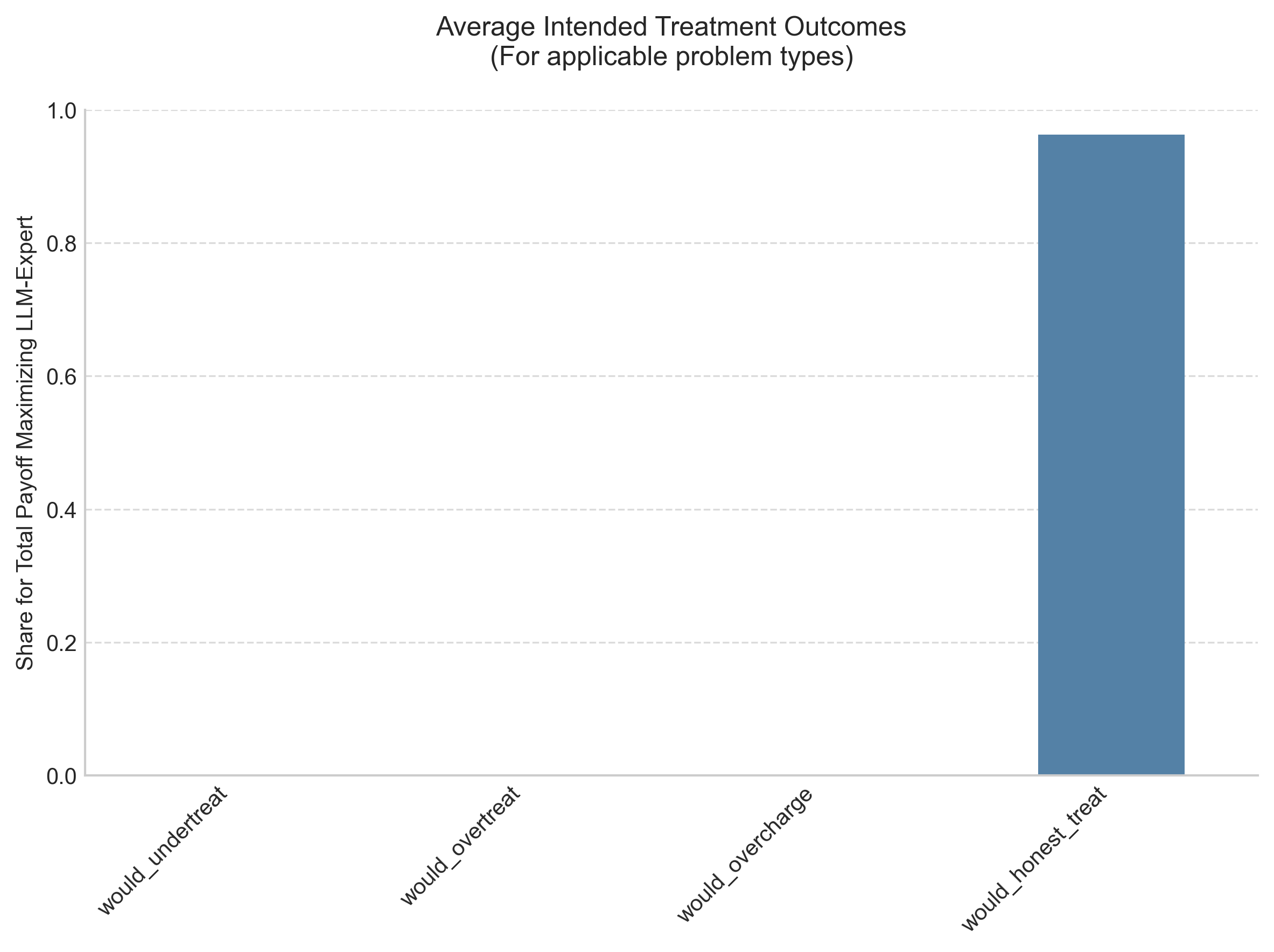}      
    \caption{LLM expert treatment and price-charging behavior in \textit{No Institution}, conditional on the LLM's objective function prompt.}
    \label{fig:aiai_noinstitution}
\end{figure}

Figure \ref{fig:llm_prices_aiai} illustrates the expected value experts offer to consumers via their chosen prices, assuming standard model expectations. While the objective function influences price-setting, it has little influence on the consumer's optimal strategy. Expected payoffs remain below the outside option, and consumers would open themselves up to exploitation by approaching an expert.

Treatment behavior (Figure \ref{fig:aiai_noinstitution}) reveals LLM choices that are closely connected to theory. Without any objective prompt, the LLM exhibits a substantial tendency to defraud their consumers, vindicating the latter's decisions to not enter the market. The self-interested AI-expert basically follows the game-theoretical predictions by always undertreating and overcharging the consumer. Changing the objective towards fairness concerns or total payoff shifts behavior towards honest treatments. This means that, while an inequity averse LLM expert does post the highest prices, they also follow up with mostly honest treatments, which would lead to an equal distribution of gains. In these cases, market efficiency could have been substantially increased if LLM consumers had some initial trust in the LLM expert's good intentions, or if they had access to the underlying objective function. Overall, we show that LLMs are able to play strategically in one-shot credence goods markets with asymmetric information, but never create a successful interaction without any market institutions. This is because LLM experts set too high prices, and LLM consumers do not trust LLM experts to not defraud them. Furthermore, the default model often opts to exploit their information advantage at the expense of consumers. LLMs can be prompted to play honestly, which has no positive effect on overall market welfare due to a lack of market institutions and objective transparency.

\subsubsection{AI-AI-Interaction -- Verifiability}

Verifiability can theoretically solve the information asymmetry problem on credence goods markets through competition and expert commitments to honesty. Prior experimental evidence suggests that, in contrast to theoretical predictions, verifiability often does not lead to sustained market interaction, potentially due to social preferences like expert inequity aversion. Our simulations confirm the same for AI-AI interactions with LLM agents. Irrespective of the expert's objective function, no LLM consumer ever approaches an expert. One reason is that LLM experts mostly do not commit to honesty, but set prices such that they have transparent incentives to always rely on the low cost treatment while charging $\ubar{p}>3$ (see Table \ref{tab:expert_prices} and Figure \ref{fig:llm_prices_aiai}), which is too high. Hence, consumers predict a 50\% chance of undertreatment and maximize their expected payoff by relying on the outside option.

\begin{figure}[t!]
    \centering
    \includegraphics[width=0.4\textwidth]{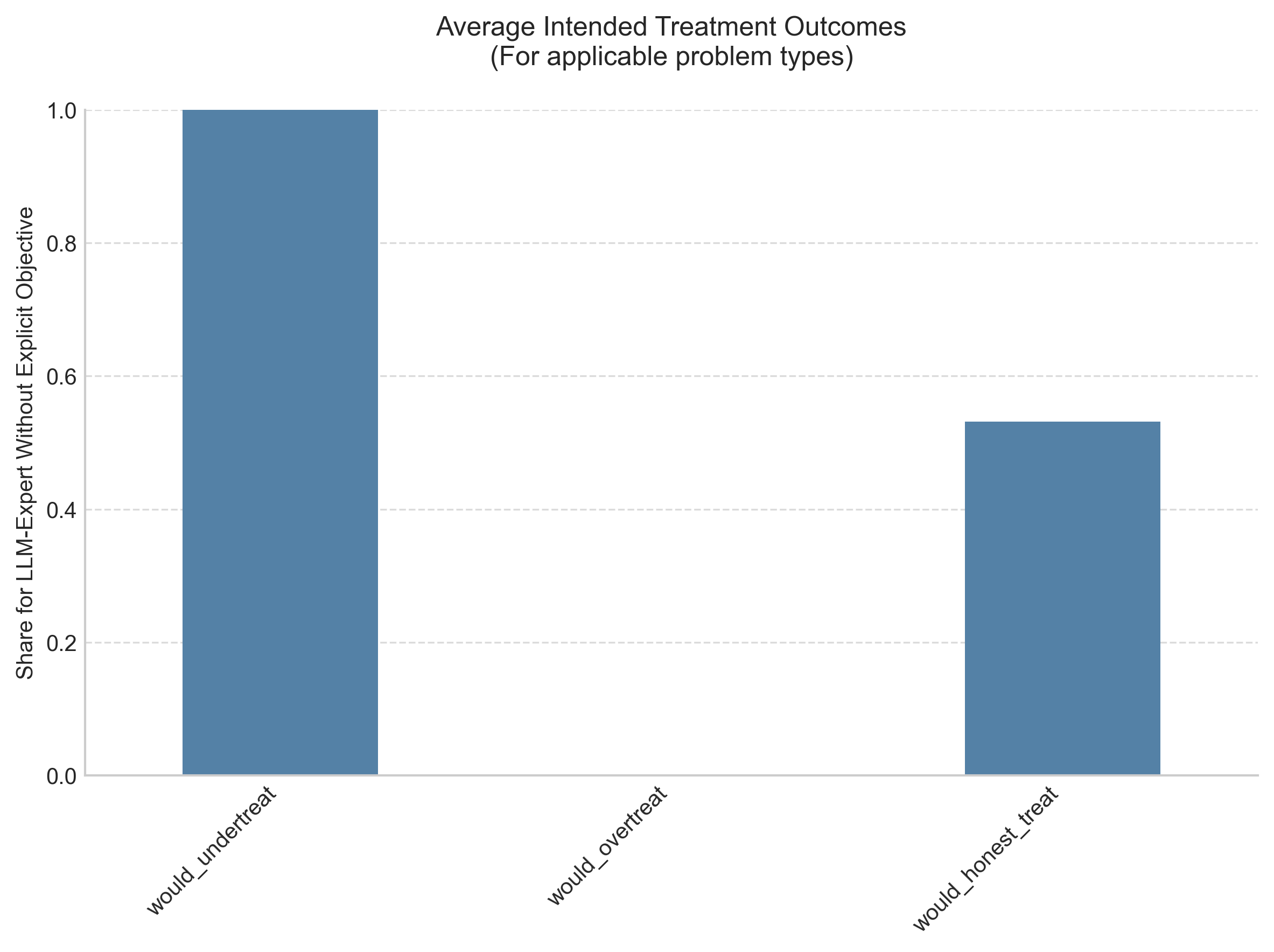}
    \hspace{0.05\textwidth}
    \includegraphics[width=0.4\textwidth]{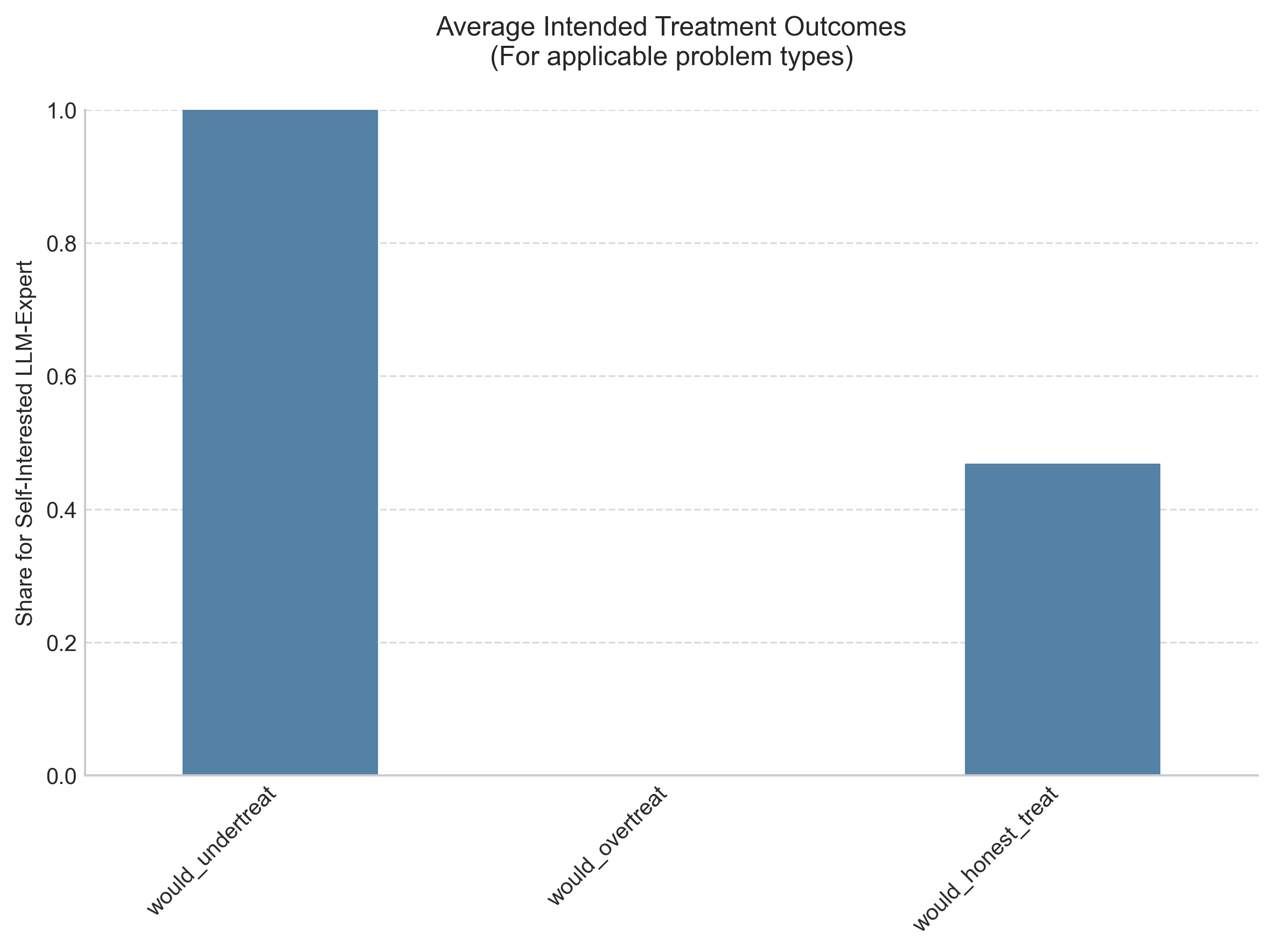} 
    \includegraphics[width=0.4\textwidth]{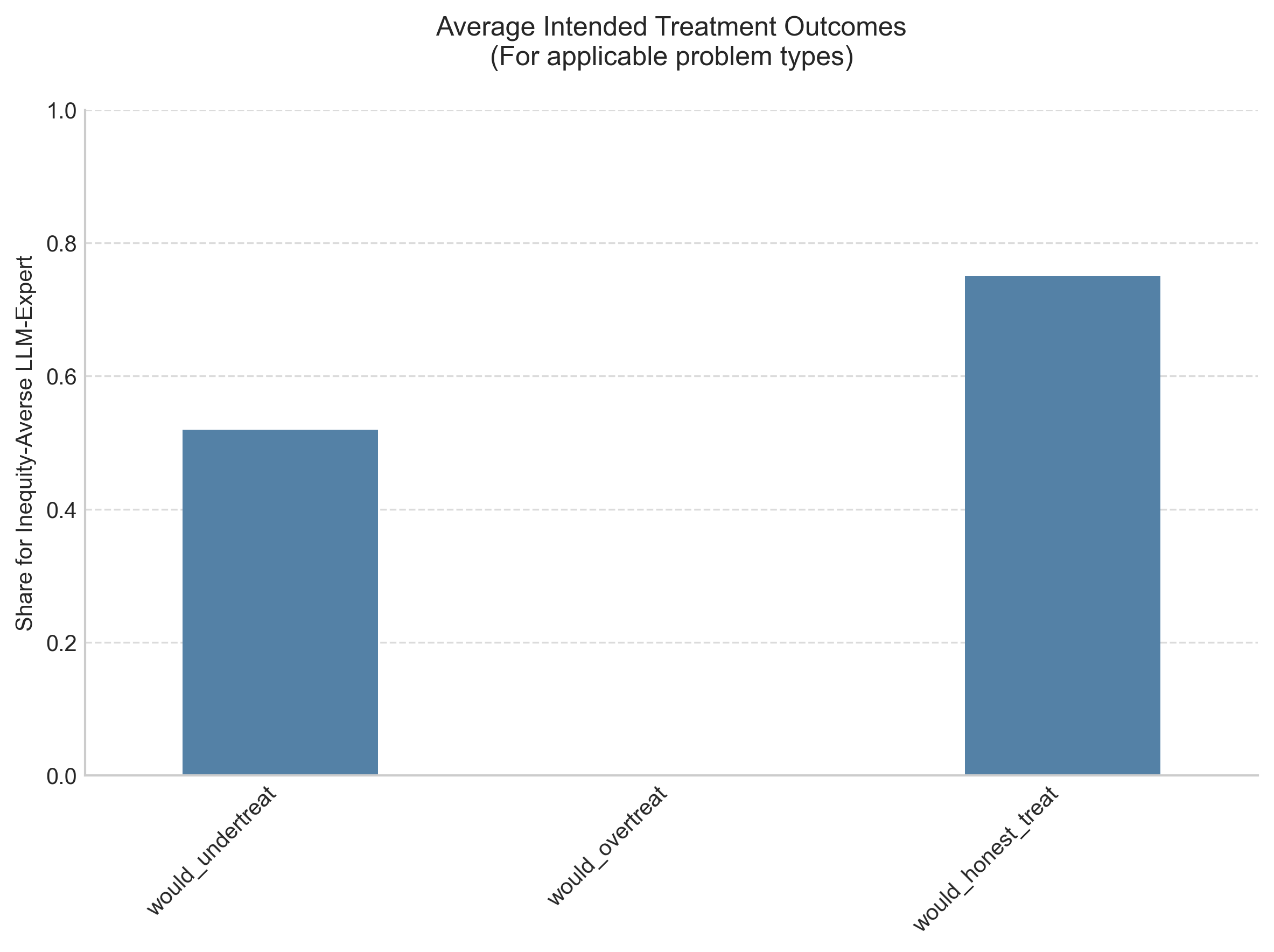}
    \hspace{0.05\textwidth}
    \includegraphics[width=0.4\textwidth]{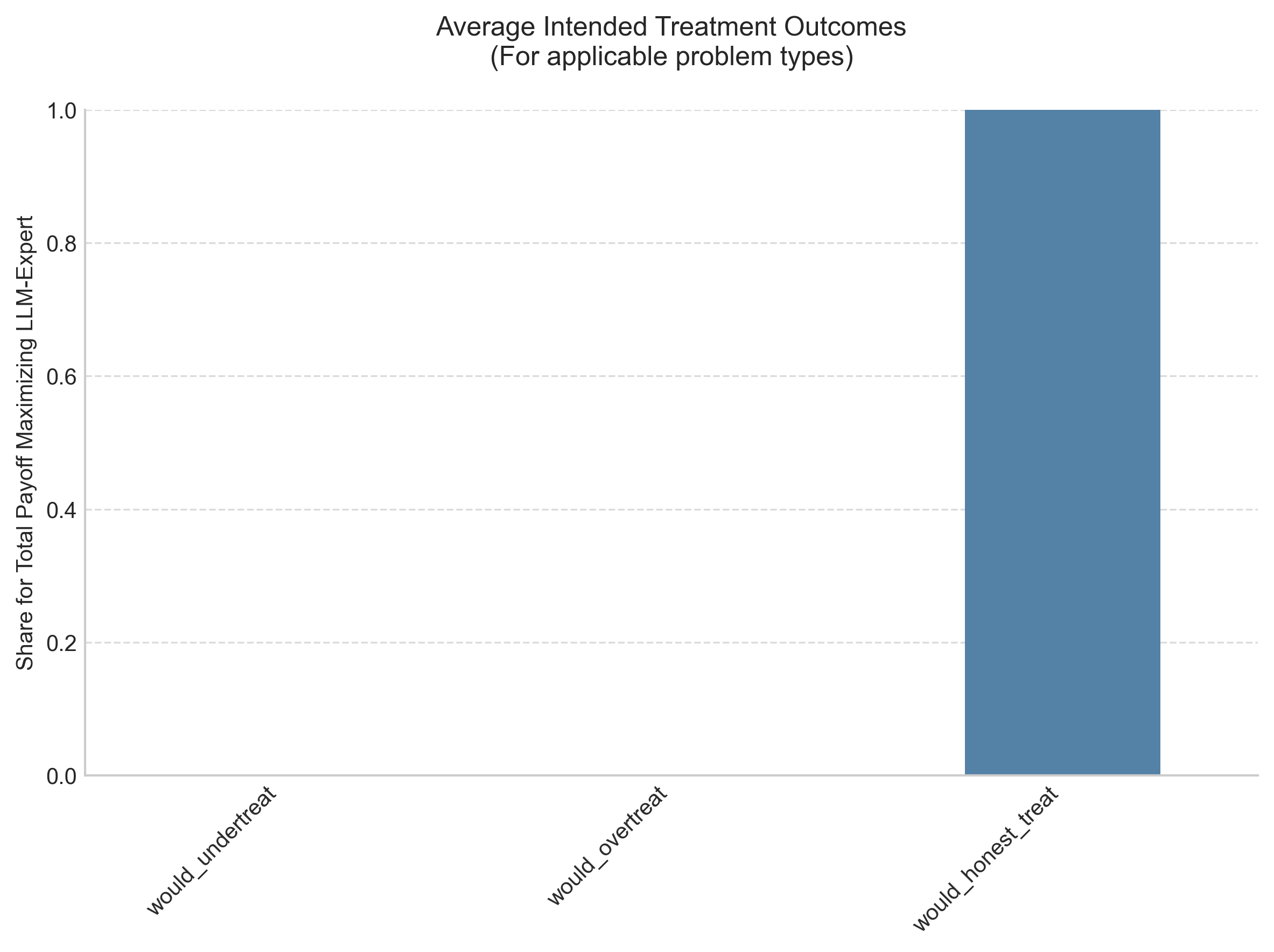}      
    \caption{LLM expert treatment and price-charging behavior in \textit{Verifiability}, conditional on the LLM's objective function prompt.}
    \label{fig:aiai_veri}
\end{figure}

However, even with the efficiency-loving LLM expert whose prices do signal indifference between the two treatments due to equal markups, LLM consumers choose to not approach the expert. As shown in Figure \ref{fig:llm_prices_aiai}, under the standard assumption that profit-indifferent experts choose honesty over dishonesty, the efficiency-loving LLM offers consumer profits close to theory that clearly beat the outside option. Yet, LLM consumers apparently do not assume that LLM experts treat them honestly without any explicit incentive to do so and hence avoid them.

Expert treatment decisions (Figure \ref{fig:aiai_veri}) show that the LLM's strategy focuses on always choosing the LCT, as suggested by their pricing behavior. Only when endowing it with explicit social preferences does honest behavior increase. Once again, the efficiency-loving LLM is -- in line with their profit motive -- the most (and only) honest LLM. The inequity-averse LLM still exhibits a considerable tendency to rely on the LCT after diagnosing a big problem. Hence, the LLM consumer's assumption that LLM experts will defraud them in the absence of clear incentives is mostly well-founded and in line with expert pricing and treatment behavior, but can also be incorrect, given the right social preferences. Like before, aggressive expert price-setting interacts with a lack of consumer trust to inhibit successful market outcomes. Competition does not appear to discipline LLM experts.

\subsubsection{AI-AI-Interaction Liability}

\begin{figure}[t]
    \centering
    \includegraphics[width=0.4\textwidth]{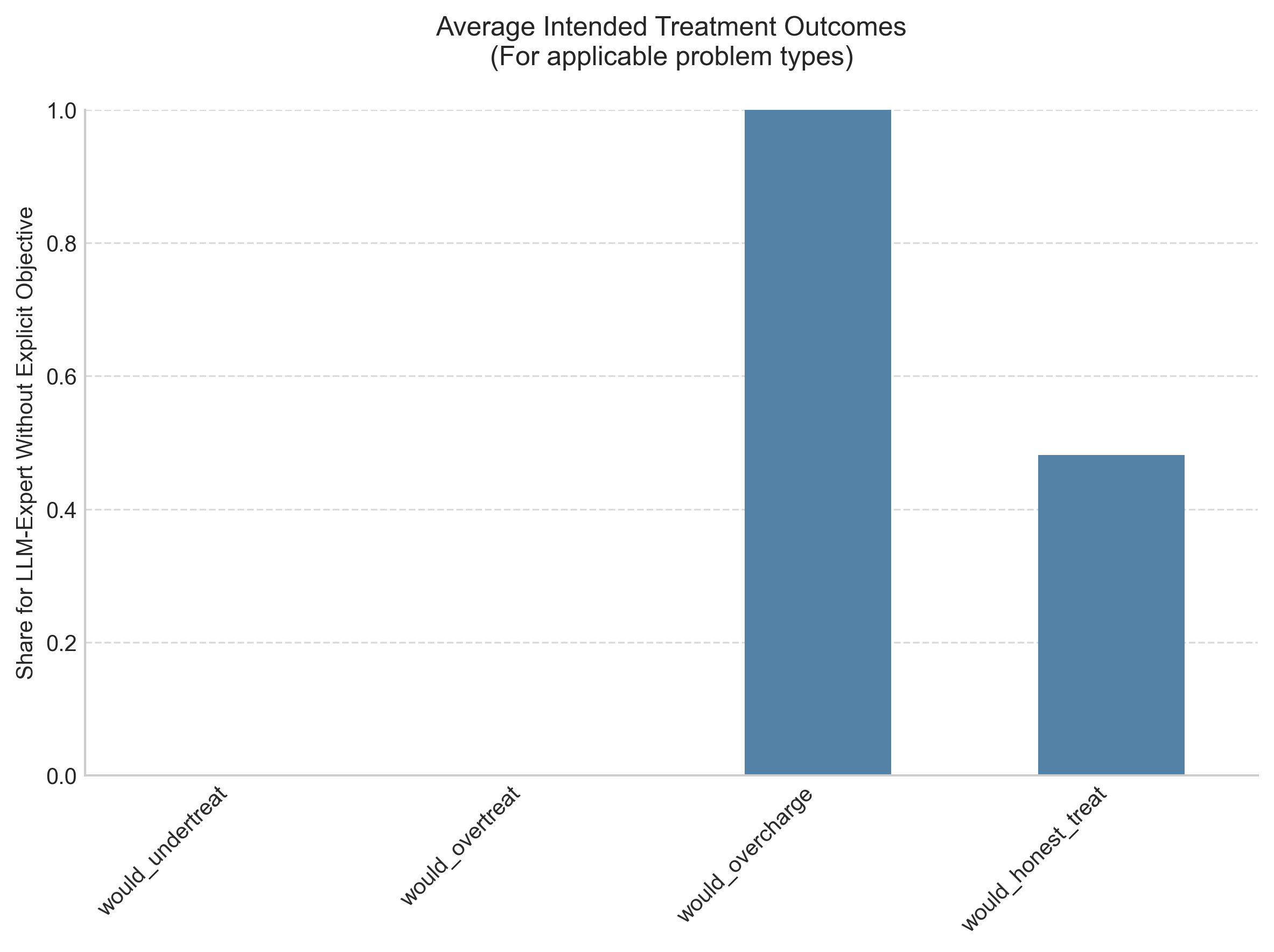}
    \hspace{0.05\textwidth}
    \includegraphics[width=0.4\textwidth]{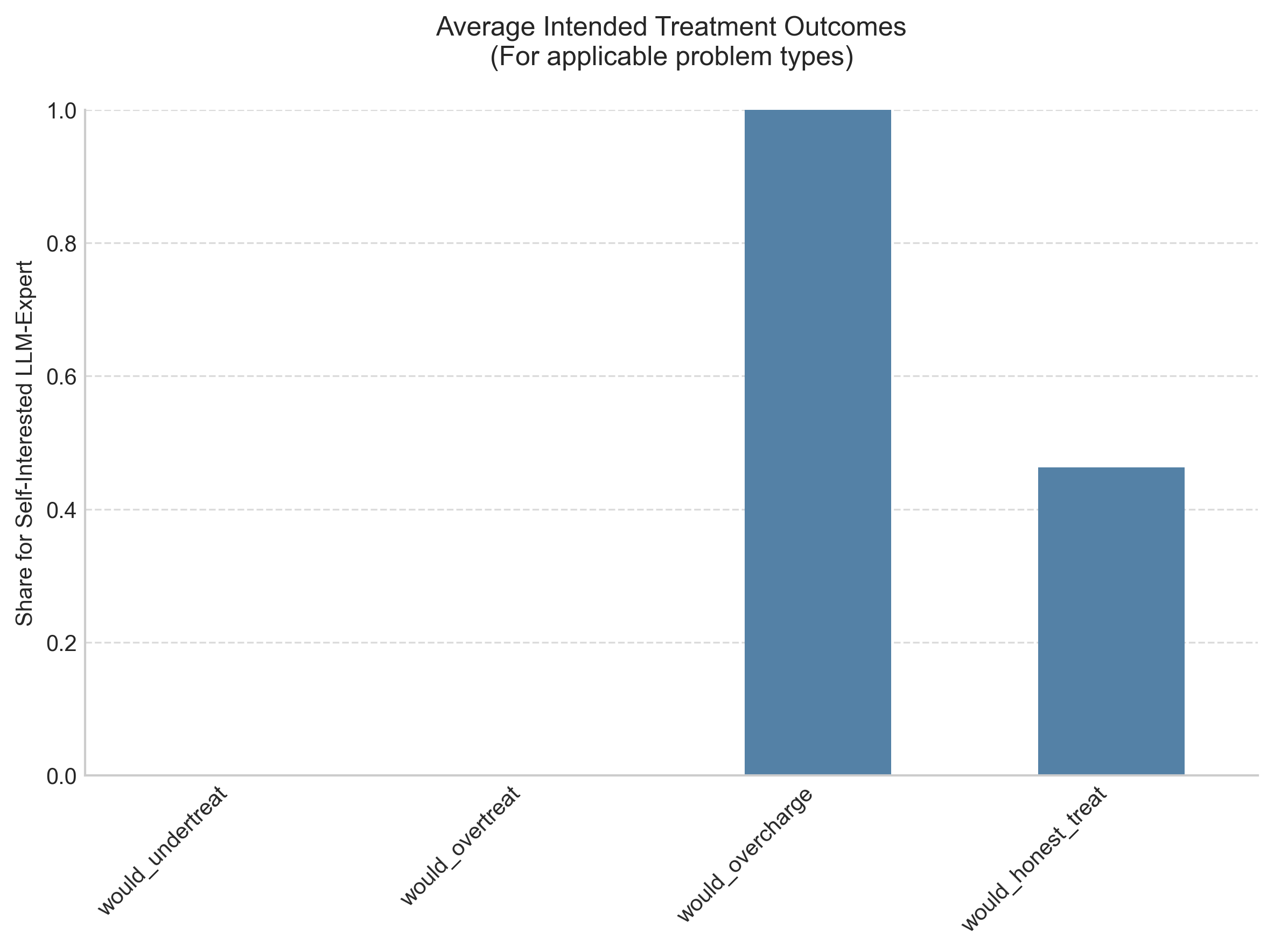} 
    \includegraphics[width=0.4\textwidth]{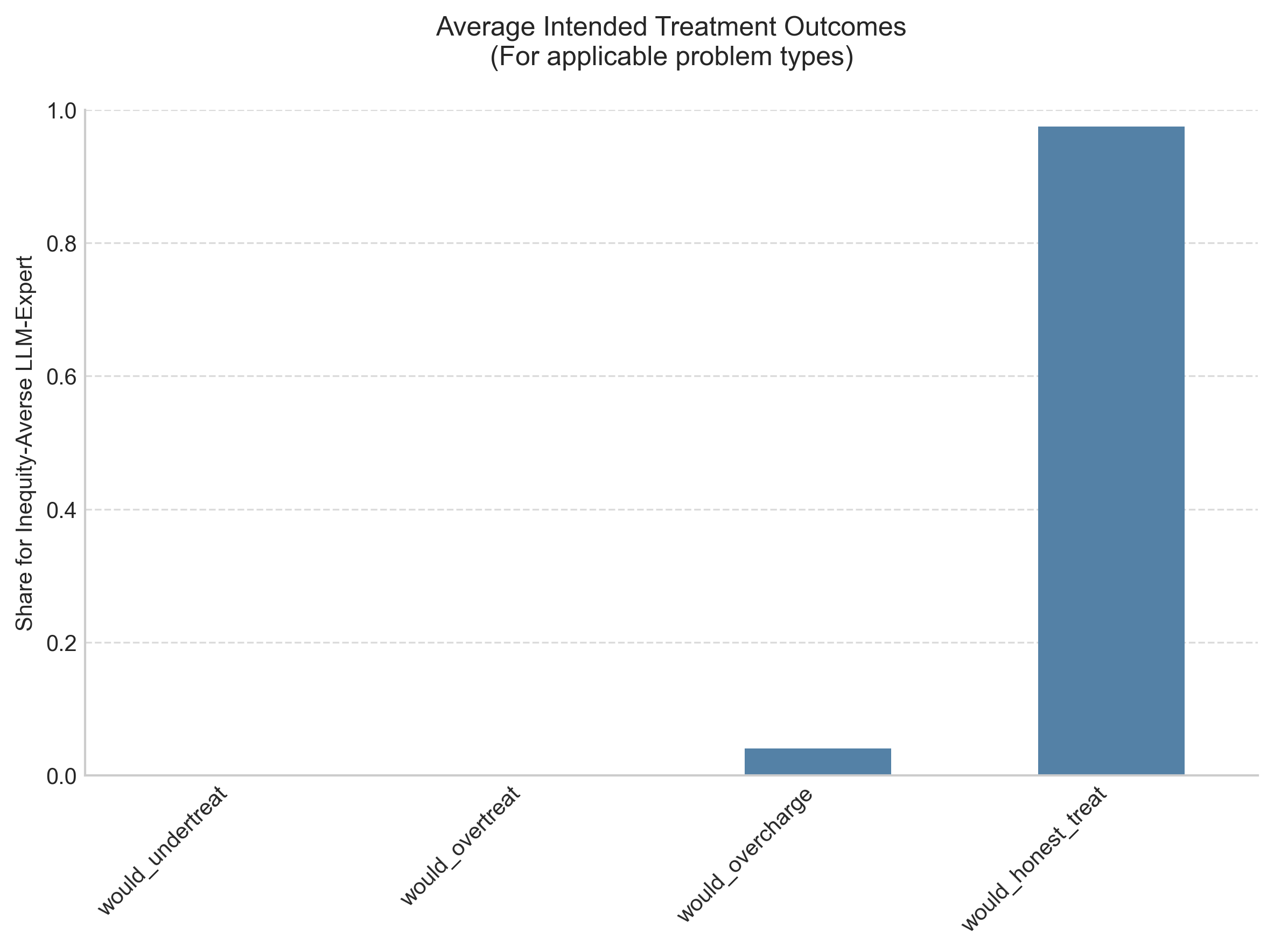}
    \hspace{0.05\textwidth}
    \includegraphics[width=0.4\textwidth]{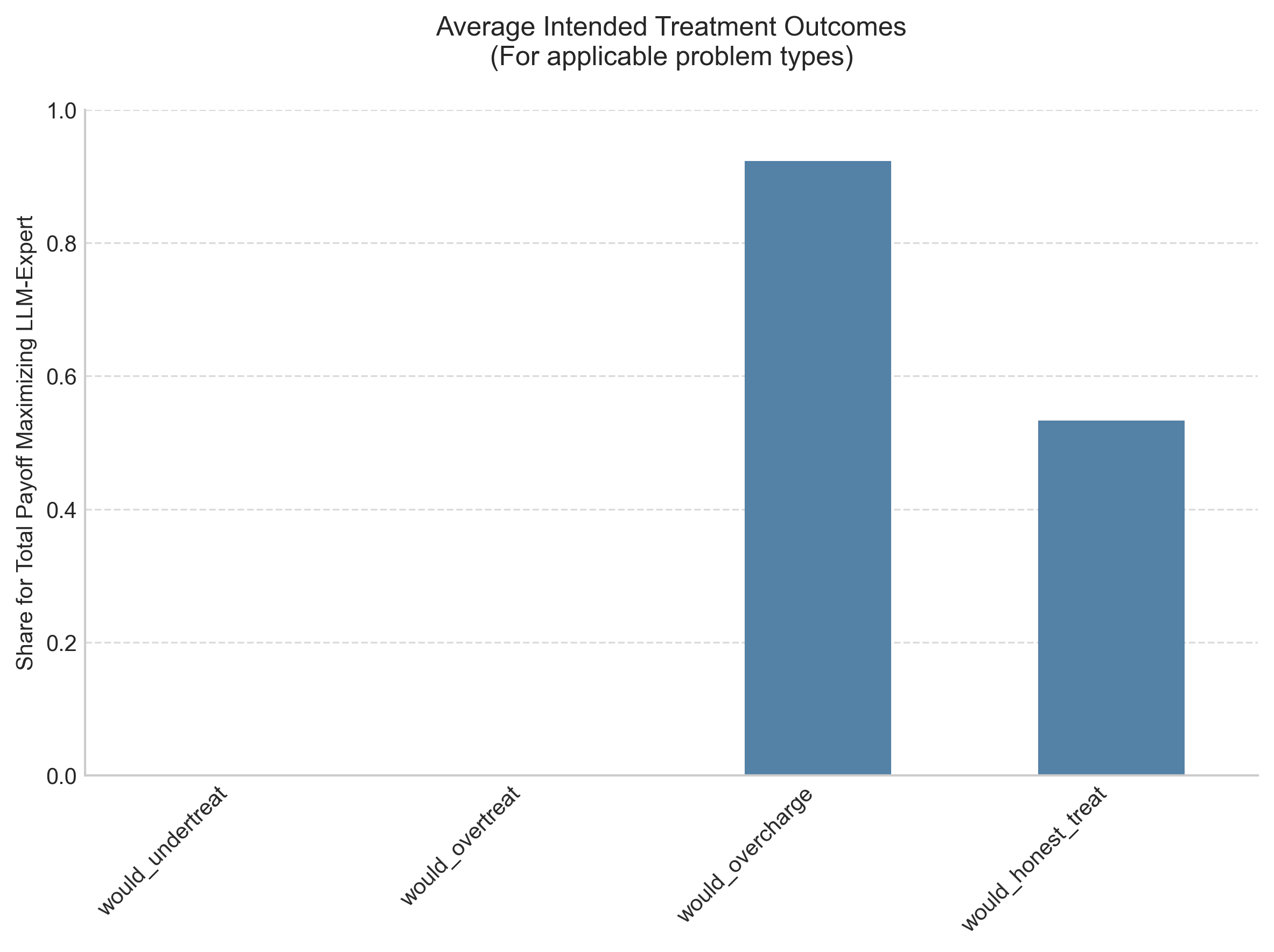}      
    \caption{LLM expert treatment and price-charging behavior in \textit{Liability}, conditional on the LLM's objective function prompt.}
    \label{fig:aiai_lia}
\end{figure}

Finally, under liability, experts must solve the consumer's problem, but are free to choose or charge any treatment and price if they diagnose a small problem. The dominant expert strategy is to always rely on $\bar{p}$. This intervention has consistently lead to high and sustained market activity on credence goods markets, both in theory and empirically. Concurrently, our simulations show that, in contrast to the prior two environments, 100\% of LLM consumers choose to approach an LLM expert.

Prices for the HCT are generally higher than predicted under competition ($\bar{p} = 5$), conforming mostly to (near) monopolist pricing ($\bar{p}_m = 8$). Still, they consistently lead to expected consumer payoffs $\pi^c_{l} > \sigma$, independent from actual expert honesty (Figure \ref{fig:llm_prices_aiai}).  Concerning treatment behavior (Figure \ref{fig:aiai_lia}), experts always opt to overcharge consumers, except if they are endowed with inequity-averse preferences. Because an efficiency-loving expert does not care about the distribution of economic gains, overcharging consumers is consistent with the objective specifically because in expectation, consumers will always approach an expert, even if they predict overcharging. Hence, while the efficiency-loving expert minimizes expert fraud under a verifiability framework, this patterns flips under liability. Here, only an inequity-averse LLM expert behaves honestly.

\subsubsection{AI-AI-Interaction -- Main Takeaways}

The simulations confirm the crucial effect of institutions and social preferences for AI-AI credence goods market outcomes. Table \ref{tab:aiai_overview} summarizes expert and consumer outcomes for each institutional and objective function condition.

\begin{table}[h!]
\centering
\caption{AI-AI Market Outcomes by Institutional Environment and Expert Objective Function}
\label{tab:aiai_overview}
\begin{tabular}{lccccccc}
\hline \hline
& & \multicolumn{4}{c}{Objective Function} \\
\cline{3-6}
Institution & Outcome & No & Self- & Inequity- & Efficiency- \\
& & Objective & Interested & Averse & Loving \\
\hline
& Share Approaching (\%) & 0 & 0 & 0 & 0 \\
& Expert Behavior  & Dishonest & Dishonest & Honest & Honest \\
No Institution & Consumer Surplus & 6.4 & 6.4 & 6.4 & 6.4 \\
& Expert Surplus & 0 & 0 & 0 & 0 \\
\addlinespace
& Share Approaching (\%) & 0 & 0 & 0 & 0 \\
& Expert Behavior  & Dishonest & Dishonest & 50/50 & Honest \\
Verifiability & Consumer Surplus & 6.4 &  6.4 & 6.4 & 6.4 \\
& Expert Surplus & 0 & 0 & 0 & 0 \\
\addlinespace
& Share Approaching (\%) & 100 & 100 & 100 & 100 \\
& Expert Behavior & Dishonest & Dishonest & Honest & Dishonest \\
Liability & Consumer Surplus & 12 & 12 & 15.68 & 8.64 \\
& Expert Surplus & 12 & 12 & 8.32 & 15.36 \\
\hline \hline
\multicolumn{6}{p{0.95\textwidth}}{\small \textit{Notes:} Consumer and Expert Surplus are always aggregated on the group level (4 players). } \\
\end{tabular}
\end{table}

First, LLM behavior on one-shot credence goods markets does not follow theoretical predictions from the standard model. In particular, LLM experts set prices that are too high, and often do not consider how their own economic incentives shape expected consumer income. This is mostly independent from induced social preferences. Furthermore, prices are generally oriented around monopolist prices, rather than competition prices. LLM consumers are risk-averse, and only approach an expert if their expected income exceed their outside option assuming narrowly self-interested expert behavior. Second, consumers only approach experts under liability. Otherwise, the market breaks down. Third, LLM expert treatment behavior depends on the LLM's objective function. Without any, or with the explicit instruction to maximize the expert's income, the LLM always behaves dishonestly and defrauds consumers whenever possible. The effect of inducing social preferences depends on the institutions. Under \textit{Verifiability}, inequity-averse LLMs continue to undertreat consumers with a probability of 50\%, whereas the efficiency-loving LLM is always honest. Under \textit{Liability}, the efficiency-loving LLM dishonestly overcharges consumers with a probability of 92\%, whereas the inequity-averse LLM charges the honest price with a probability of 98\%.

\subsection{Results -- Human Human Interaction}
\begin{figure}[h]
    \centering
    \includegraphics[width=0.49\textwidth]{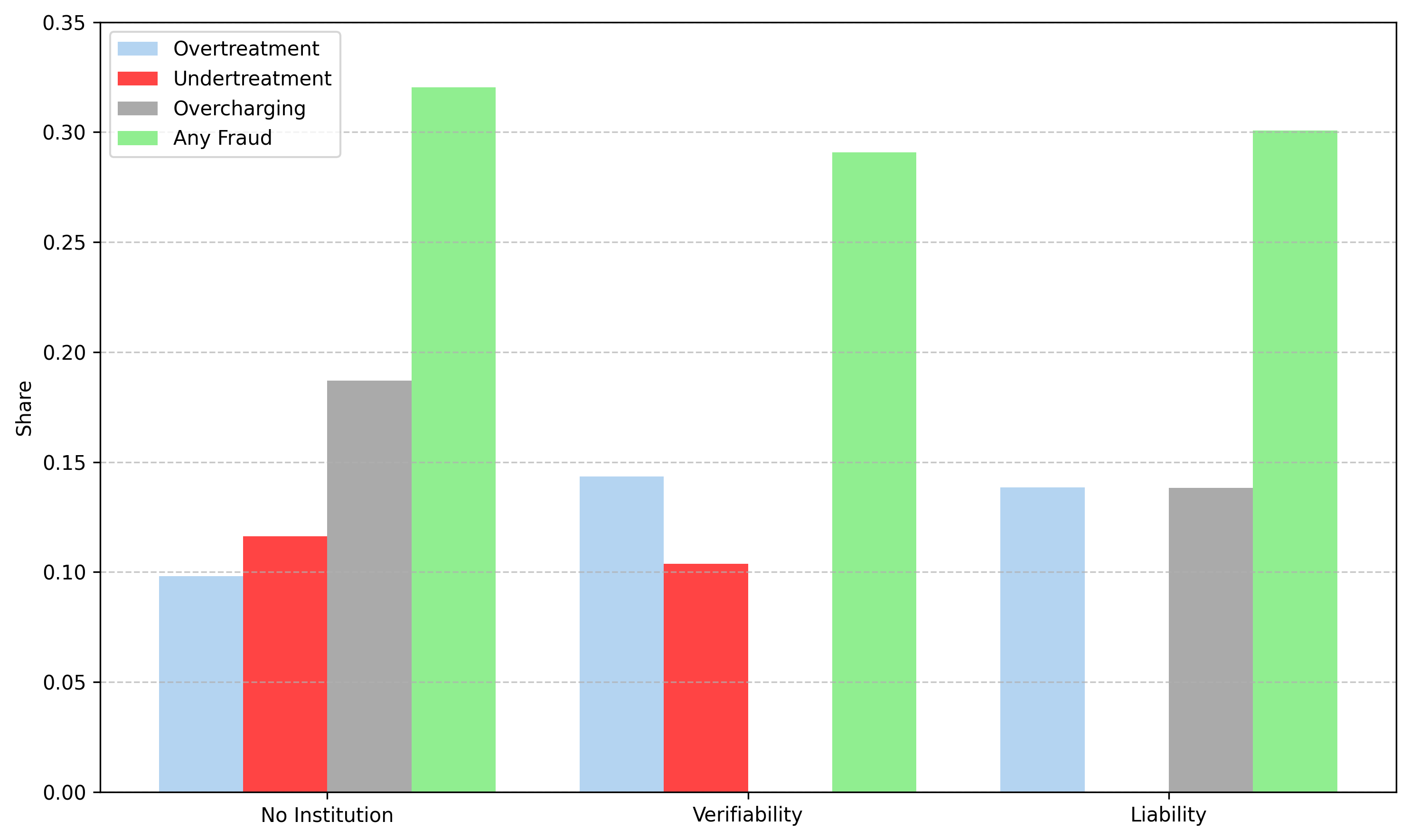}
    \includegraphics[width=0.49\textwidth]{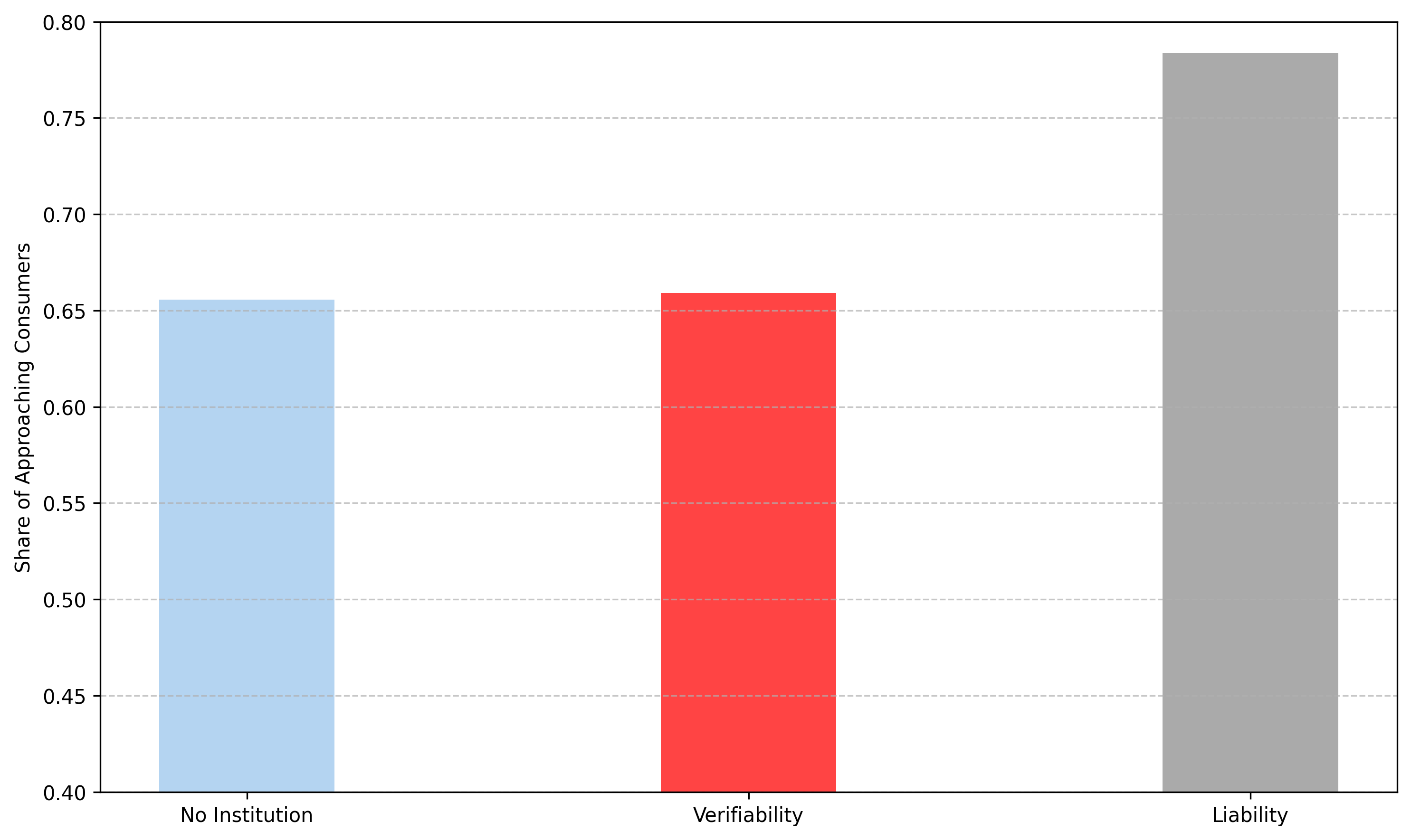}
    \caption{Left: Intended Expert Fraud Shares. Right: Consumer Approach Decisions Across the three Institutional Frameworks.}
    \label{fig:approach_treatment_hh}
\end{figure}

\begin{figure}[h]
    \centering
    \caption{Human Expert Price Setting and Expected Consumer Payoffs For the Nine Most Common Price Pairs. Large symbols represent higher frequencies.}
    \includegraphics[width=0.9\textwidth]{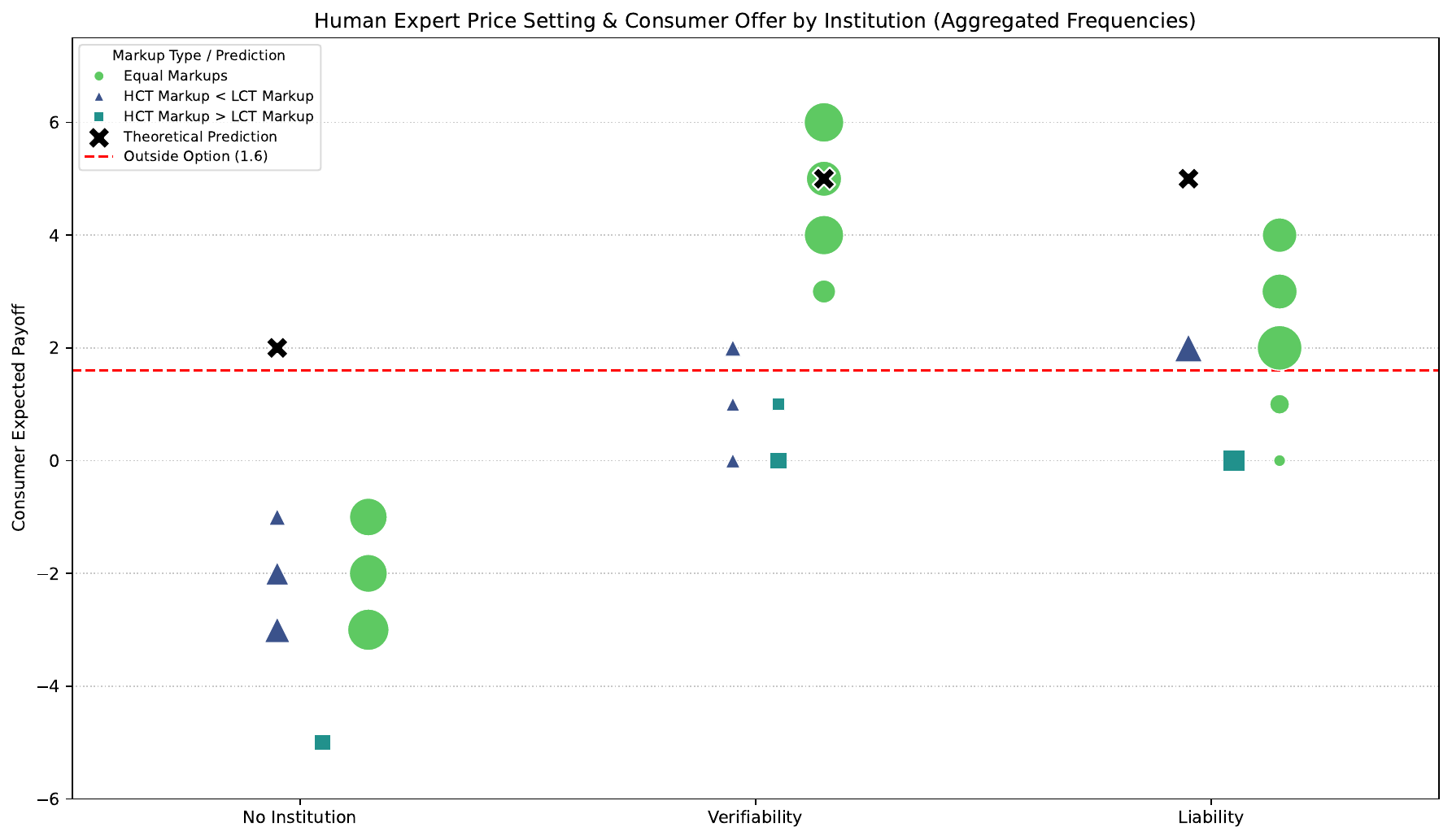}
    \label{fig:llm_prices_humanhuman}
\end{figure}

How do AI markets differ from human markets? First, we look at expert behavior. Table \ref{tab:price_pairs_rankings} and Figure \ref{fig:llm_prices_humanhuman} show human expert price-setting. There are two substantial differences. One, human experts are more likely to rely on equal markups, and two, except for the efficiency-loving LLM, prices in \textit{Verifiability} are much more aligned with theory, leading to higher expected consumer payoffs. Treatment behavior confirms that fraud exists, but is relatively rare. In particular, only between 15 and 20\% of experts choose to overcharge consumers, despite having strong economic incentives to do so (Figure \ref{fig:approach_treatment_hh}).

\begin{table}[h]
\centering
\caption{Consumer Approach Behavior: Treatment Effects and Descriptives}
\label{tab:cons_approach}
\begin{tabular}{lccccc}
\hline \hline
& \multicolumn{2}{c}{Repeated Measures Logit} & \multicolumn{3}{c}{Between-Subject Logit} \\
& (1) & (2) & (3) & (4) & (5) \\
& GEE & GEE with Controls & No Institution & Verifiability & Liability \\
\hline
Verifiability & 0.003 & 0.003 & & & \\
& (0.025) & (0.027) & & & \\
Liability & 0.128$^{***}$ & 0.127$^{***}$ & & & \\
& (0.027) & (0.025) & & & \\
Risk & & 0.071$^{***}$ & 0.077$^{***}$ & 0.079$^{***}$ & 0.054$^{***}$ \\
& & (0.007) & (0.008) & (0.008) & (0.009) \\
Age & & -0.004$^{**}$ & -0.005$^{*}$ & -0.005$^{*}$ & -0.001 \\
& & (0.001) & (0.002) & (0.002) & (0.002) \\
Female & & 0.078$^{*}$ & 0.133$^{**}$ & 0.063 & 0.038 \\
& & (0.035) & (0.048) & (0.049) & (0.045) \\
\hline
Observations & 915 & 915 & 305 & 305 & 305 \\
\hline \hline
\multicolumn{6}{l}{\footnotesize Notes: Columns 1-2 show population-averaged effects from GEE with unstructured correlation.}\\
\multicolumn{6}{l}{\footnotesize Dependent variable: binary variable capturing whether a consumer chose to approach an expert.} \\
\multicolumn{6}{l}{\footnotesize Columns 3-5 show average marginal effects from separate logistic regressions.} \\
\multicolumn{6}{l}{\footnotesize Standard errors in parentheses. $^{*}$ p$<$0.05, $^{**}$ p$<$0.01, $^{***}$ p$<$0.001} \\
\end{tabular}
\end{table}

In an anonymous one-shot environment, there are no future reputation costs, and consumers decide whether to approach an expert before learning about their outcome. Still, most experts charge prices in line with their prior treatment decision. Because each expert makes four treatment decisions, these numbers under-estimate the share of experts who are \textit{sometimes} dishonest. Irrespective of the institutional environment, roughly 30\% of experts intend to defraud at least one consumer. Hence, while the majority is always honest, a sizeable minority does engage in fraudulent treatment behavior.

Compared to the AI interactions above, human experts are much more honest than the self-interested LLM or the LLM without an explicit objective. This is in line with human experts pricing behavior, which rarely adjusts to potential consumer exploitation. However, endowing LLMs either with (1) inequity-averse or (2) efficiency-loving social preferences can, depending on the institutional environment, exceed human experts in terms of honesty. Crucially, without any institutions, both the inequity-averse and the efficiency-loving LLM exhibit more honest treatment behavior than human experts. Under verifiability, this \textit{only} holds for the efficiency-loving LLM, whereas under liability, this \textit{only} holds for the inequity-averse LLM.

\noindent
\textbf{Consumer Behavior.} Figure \ref{fig:approach_treatment_hh} depicts consumer approach shares across the three institutional frameworks. In both \textit{No Institution} and \textit{Verifiability}, roughly 35\% of consumers choose to leave the market and rely on their outside option. For the free market setup, this is substantially better than -- given prices -- theory would predict, and in line with established experimental evidence. Like in prior multi-round setups, verifiability does not improve market efficiency. This is despite theoretically sound expert pricing that generally does not incentivize dishonest treatment behavior. Participation is therefore lower than predicted by the standard model, but much higher than in the AI-AI simulations. Finally, a liability institution substantially increases market participation to 80\%. Table \ref{tab:cons_approach} shows repeated measures logistic regressions to test for concurrent within-subject treatment effects on consumer approach behavior (columns 1 \& 2), as well as standard cross-sectional logit regressions for some demographic variables on approach behavior, conditional on the institutional environment. First, we see that the positive effect of liability on successful market interactions is large and significant. Second, general consumer approach behavior correlates with risk-preferences and participant sex. Specifically, without any market institutions, where experts are free to thoroughly exploit consumers by choosing the cheapest treatment while charging the highest price, both, stronger risk-preferences as well as being female significantly predict consumer-expert interactions. The inclusion of market institutions alleviates gendered-effects, whereas risk preferences remain robust predictors of consumer behavior.

\noindent
\textbf{Market Efficiency and Surplus Distribution.} Finally, we look at market outcomes and efficiency estimations.\footnote{We exclude 3 consumer observations because of a price display error in the liability section of the experiment. Due to the random assignment of participants and groups within treatments during the survey, group size is not perfectly evenly distributed. We account for group size differences in all analyses.} Table \ref{tab:efficiency_hh} displays efficiency ratios as well as role-specific average group surplus across the three institution treatments. Throughout, consumers achieve a substantially larger surplus than experts, which can be explained by low prices and relatively low rates of expert fraud, specifically under-treatment. As expected, markets with a liability rule are significantly more efficient than free ($t = 5.01, \, p < 0.001$) and verifiable ($ t = 4.04, \, p < 0.001$) markets. There are no differences between \textit{Verifiability} and \textit{No Institution}.

\begin{table}[h]
    \centering
    \caption{Estimated Market Efficiency, Consumer Surplus and Expert Surplus}
    \begin{tabular}{lcccc}
        \toprule
        Treatment & Relative Efficiency & Consumer Surplus & Expert Surplus & $\Delta$ \\
        \midrule
        No Institution & 0.61 & 7.4 & 3.2 & 4.2 \\
        Verifiability & 0.65 & 8.6 & 3.2 & 5.4 \\
        Liability & 0.84 & 10.1 & 5.2 & 4.9 \\
        \bottomrule
    \end{tabular}
    \vspace{0.1cm}
    \caption*{\small \textit{Note:} Group-level summary statistics. \textit{Relative Efficiency} is calculated as the ratio of actual group income and maximum potential group income. All values are derived from group averages. \textit{Consumer Surplus} denotes the average cumulative income of all consumers in a group, \textit{Expert Surplus} denotes the average cumulative income of all experts in a group.}
    \label{tab:efficiency_hh}
\end{table}

Consumers always earn more on average than experts, although the gap is relatively smaller when experts are liable. The only significant differences for consumer surplus is between \textit{Liability} and \textit{No Institution} ($t = 2.98, \, p = 0.002$). Experts benefit disproportionately from the liability rule, as their average income jumps by roughly 60\% compared to both other regimes ($t = 3.90, \, p < 0.001$; $\, t = 3.62, \, p < 0.001$). Thus, while consumers do benefit from the institutional intervention, in our setting with relatively high shares of honest treatment behavior, experts are the main beneficiaries. In contrast, verifiability has no meaningful effect on any metric, and, if anything, increases the income gap between experts and consumers. Overall, compared with the AI-AI interactions, humans drastically exceed LLM agents in the absence of a liability rule. While for the latter, the AI market reaches full efficiency, these gains primarily accrue to experts, and they cannot compensate for efficiency losses in the other two institutional cases. In so far as expert market clearing is conditional on trust between actors, humans exhibit substantial advantages over generative AI.

\subsection{Results -- Human-AI Interaction}
\begin{figure}[h]
    \centering
    \caption{LLM Expert Price Setting and Expected Consumer Payoffs in the Human-AI Simulations.}
    \includegraphics[width=0.9\textwidth]{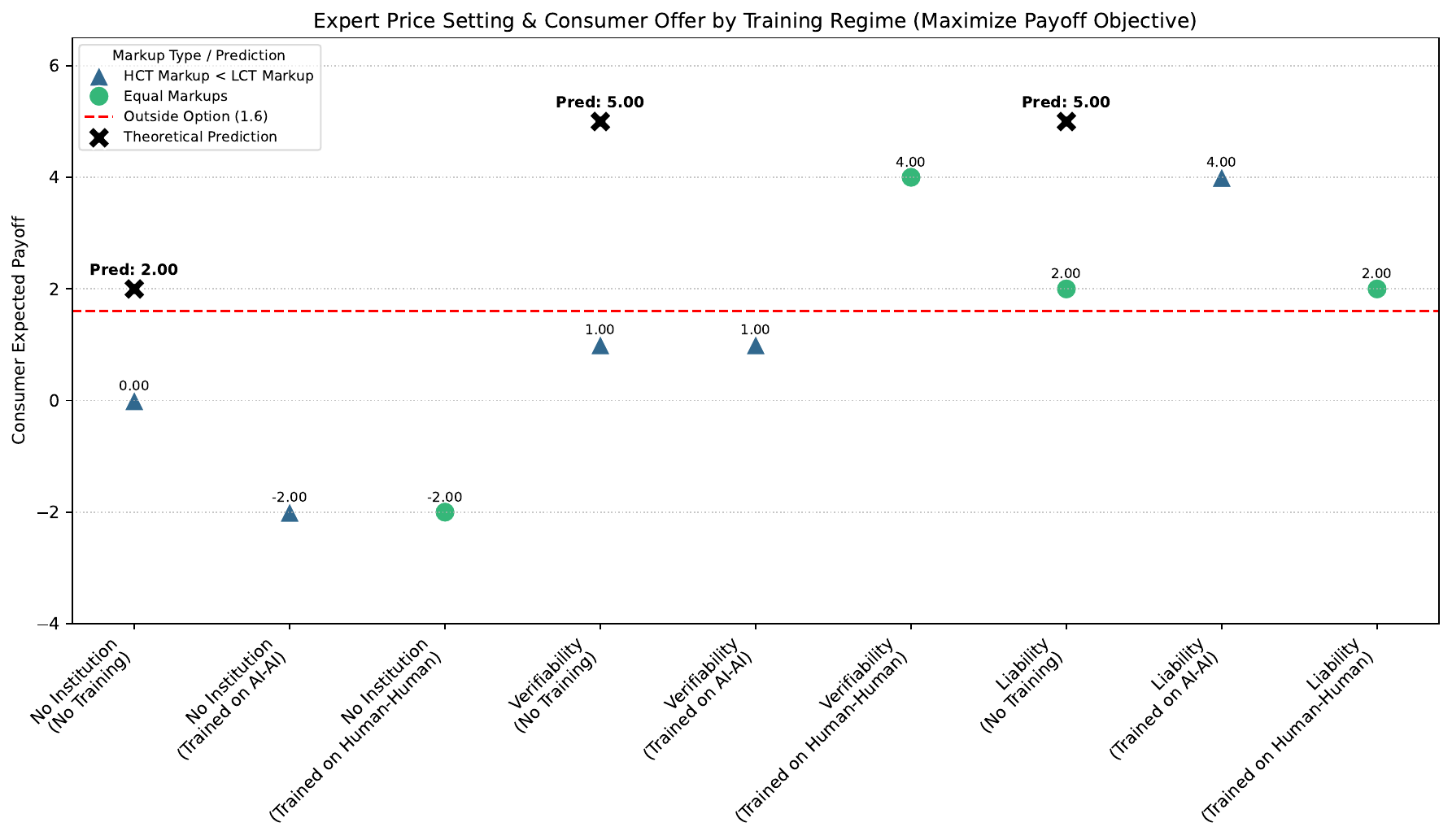}
    \label{fig:llm_prices_humanai}
\end{figure}

Now we look at the results from Human-AI Interactions where human consumers can choose to approach AI experts. Table \ref{tab:llm_expertdecisions_aiai} shows price-setting (see also Figure \ref{fig:llm_prices_humanai}), treatment and price-charging decision from the LLM. In \textbf{No Training}, the LLM only learns from instructions. In \textbf{AI Trained}, the LLM additionally receives data from 300 AI-AI simulations (see Section 2.1). In \textbf{Human Trained} the LLM receives data from 300 experimental Human-Human observations (see Section 5.2).

\begin{table}[t]
    \centering
    \caption{LLM Expert Decision Making by Treatment and Institutional Environment}
    \begin{tabular}{lccccc}
        \toprule
        Condition & Institutions & $\ubar{p}$ & $\bar{p}$ &  Treatment & Price Charged  \\
        \midrule
        \multicolumn{6}{l}{\textbf{No Training}} \\
        & No Institution    & 3.0 & 5.0 & LCT & $\bar{p}$  \\
        & Verifiability     & 4.0 & 7.0 & LCT & $\ubar{p}$  \\
        & Liability         & 4.0 & 8.0 & Honest & $\bar{p}$ \\
        \addlinespace
        \multicolumn{6}{l}{\textbf{Trained on AI-AI}} \\
        & No Institution    & 4.0 & 7.0 & LCT & $\bar{p}$  \\
        & Verifiability     & 4.0 & 4.0 & LCT & $\ubar{p}$  \\
        & Liability         & 3.0 & 6.0 & Honest & $\bar{p}$ \\
        \addlinespace
        \multicolumn{6}{l}{\textbf{Trained on Human-Human}} \\
        & No Institution    & 3.0 & 7.0 & LCT & $\bar{p}$  \\
        & Verifiability     & 4.0 & 8.0 & LCT & $\ubar{p}$  \\
        & Liability         & 4.0 & 8.0 & Honest & $\bar{p}$ \\
        \bottomrule
    \end{tabular}
    \vspace{0.1cm}
    \caption*{\small \textit{Note:} Model: claude-3-5-sonnet-20241022. The model was prompted to only care about maximizing the expert's payoff.}
    \label{tab:llm_expertdecisions_aiai}
\end{table}

First, the self-interested LLM expert basically makes the theoretically predicted treatment choices. It always chooses the LCT while overcharging consumers in the free market, generally chooses higher markups for the LCT and relies on it under verifiability, and always charges the price for the HCT with liability. As observed above, LLM experts consistently under-treat consumers under equal markups (\textbf{Human Trained}) despite no economic incentives to do so, even while interacting with humans who experience real economic consequences. Looking at price-setting (Figure \ref{fig:llm_prices_humanai}), LLMs generate offers that are similar to the AI-AI market. Only the human training data in \textit{Verifiability} causes a relevant shift for consumers' best strategy. In most cases, prices remain higher than theoretically predicted, regularly signal incentives for under-treatment, and imply an expected consumer payoff below the outside option. Specifically in \textbf{Human Trained}, there is a strong alignment with the human tendency to set prices around equal markups.

\begin{figure}[t]
    \centering
    \includegraphics[width=0.49\textwidth]{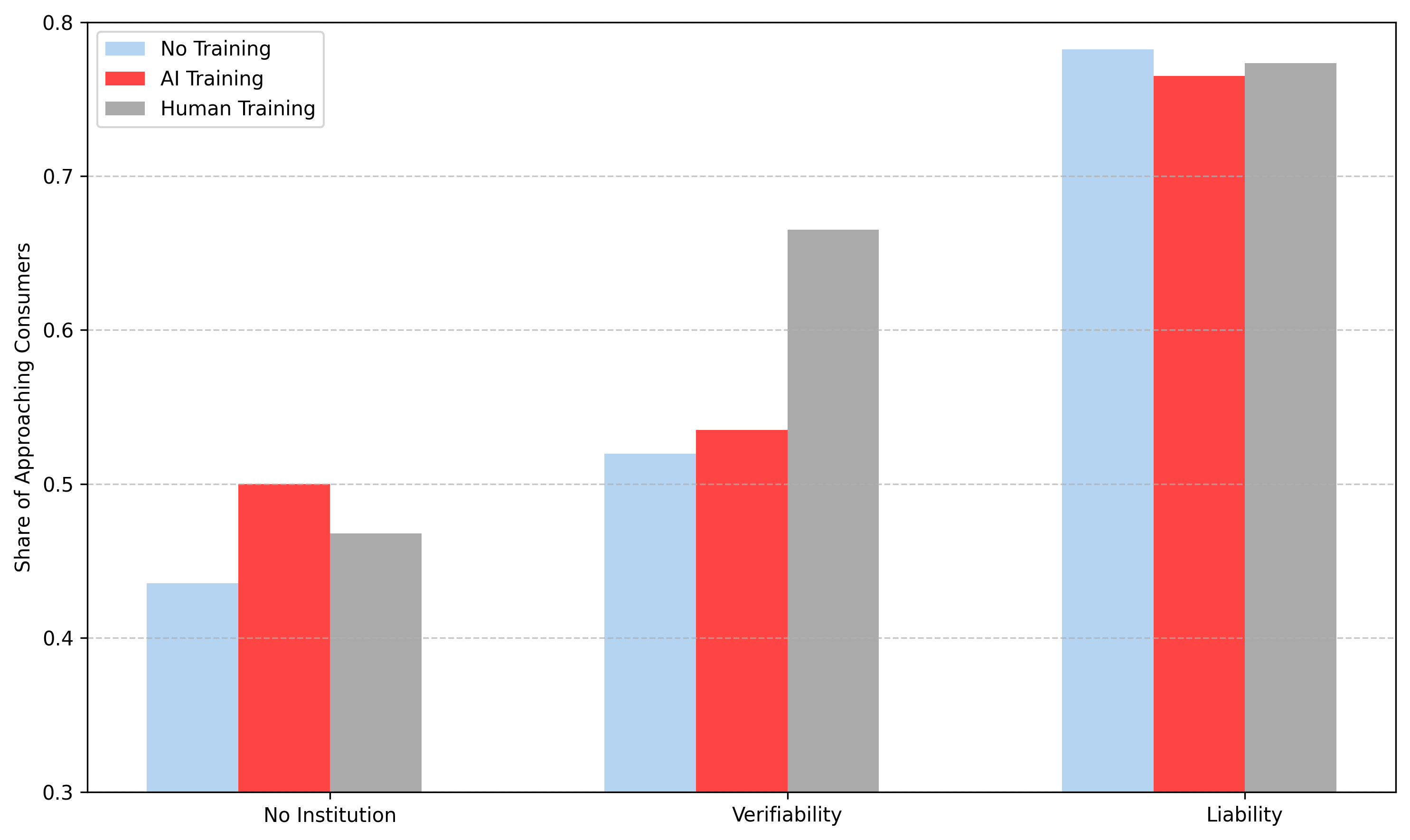}
    \includegraphics[width=0.49\textwidth]{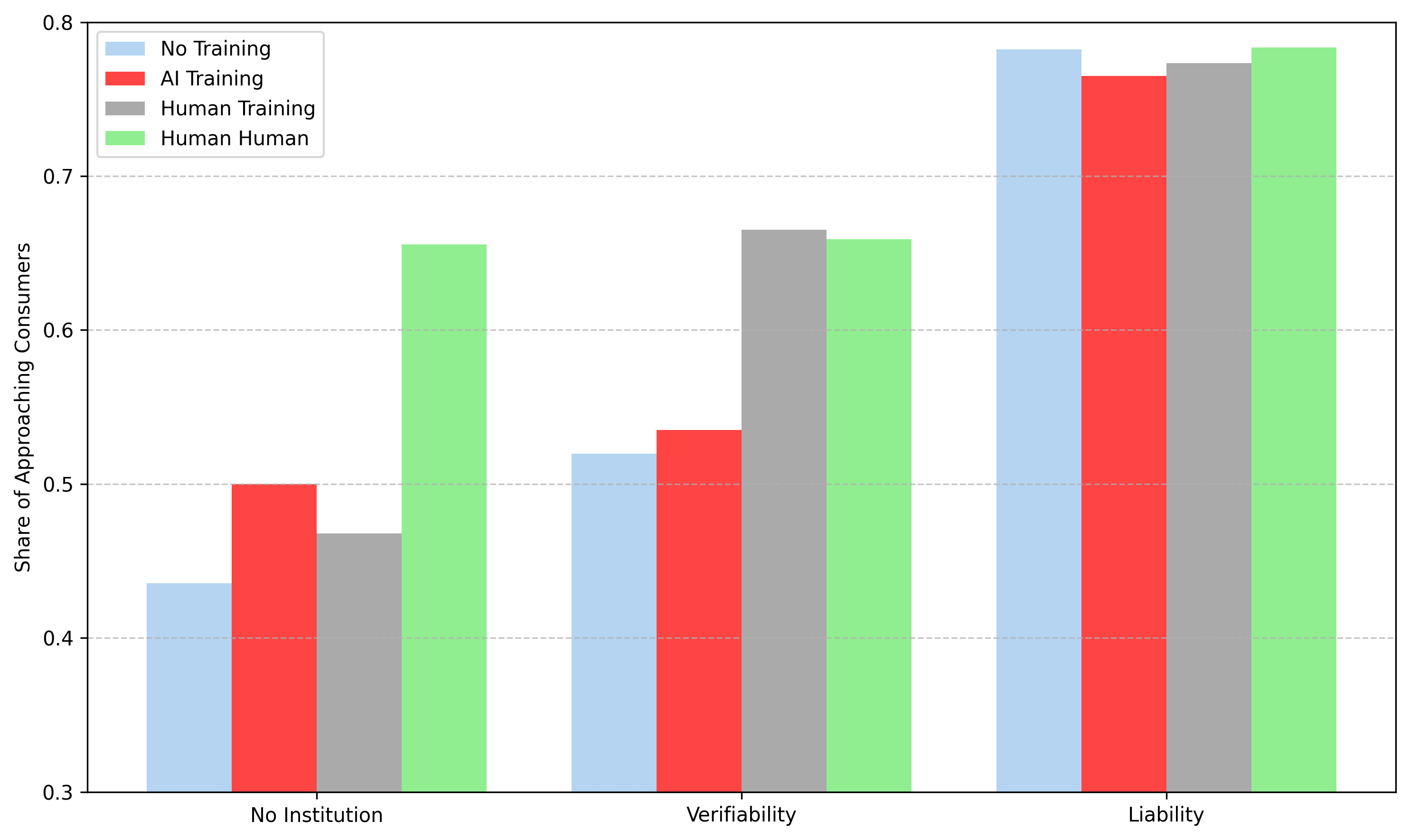}
    \caption{Consumer Approach Decisions across Institutional Frameworks}
    \label{fig:approach_treatment_hai_hh}
\end{figure}

\noindent
\textbf{Consumer Behavior.} How do consumers react to LLMs on credence goods markets? Figure \ref{fig:approach_treatment_hai_hh} shows consumer approach shares across the three institutional environments, training regimes and (right side) compared with the shares from the Human-Human experiment.
In general, patterns mirror the ones from the Human-Human environment. Consumers are much more likely to leave the market with either no institutions or a verifiability rule. When experts are liable, participation increases substantially. There are no differences between any of the training conditions for \textit{No Institution} or \textit{Liability}. Under \textit{Verifiability}, there are significant differences ($\tilde{\chi}^2 = 10.55, p = 0.005$), as consumers are much more likely to approach the LLM expert if it has been priorly trained on human data. This holds in comparison with both \textit{No Training} ($\tilde{\chi}^2 = 8.84, p = 0.003$) and \textit{AI Training} ($\tilde{\chi}^2 = 7.09, p = 0.008$). This result coincides with theoretical predictions. The LLM expert that learns from Human-Human observations sets prices with equal markups, which signals human consumers that the LLM has no profit motive to defraud them. Hence, consumers who assume an honesty preference under indifferent economic incentives -- which is true for human experts but not the self-interested LLM expert -- should approach an expert. Note that, since the self-interested LLM does not consider consumer welfare, increased consumer participation will decrease average consumer surplus (Table \ref{tab:llm_expertdecisions_aiai}). We will come back to the distributive implications of LLMs further below.

\noindent
\textbf{Human-Human vs. Human-AI Consumer Behavior.} Figure \ref{fig:approach_treatment_hai_hh} compares consumer approach shares between the three Human-AI and the Human-Human condition. First, consumers are significantly more likely to approach an expert without any institutional rules if the expert is a human (0.44 vs. 0.5 vs. 0.47 vs. 0.66, $\, \tilde{\chi}^2 = 30.44, p < 0.001$). Second, consumers are significantly more likely to approach a human expert than an LLM expert under a verifiability rule, except when the LLM has been trained on human data (0.66 vs. 0.67, $\, \tilde{\chi}^2 = 0.02, p = 0.89$). Note that here, in terms of price-setting, the only difference between the human-trained and the untrained LLM is that the human-trained LLM \textit{increases} their price for the HCT by 1, thereby ensuring markup parity. Third, if experts are liable, there are no treatment differences in consumer behavior (0.78 vs. 0.77 vs. 0.77 vs. 0.78). These results suggest that the effect of generative AI on credence goods market efficiency via the number of successful interactions is strongly contingent on both the institutional framework, as well as the LLM's training data. We find that, when consumers are fully vulnerable to exploitation, they are substantially more likely to approach a human expert. 
\begin{table}[t]
    \centering
    \caption{Human-AI: Estimated Market Efficiency, Consumer Surplus and Expert Surplus}
    \begin{tabular}{lccccc}
        \toprule
        Condition & Institutions & Rel. Efficiency & Cons. Surplus & Expert Surplus & $\Delta$  \\
        \midrule
        \multicolumn{6}{l}{\textbf{No Training}} \\
        & No Institution    & 0.37 & 3.58 & 5.19 & -1.61  \\
        & Verifiability     & 0.39 & 5.17 & 4.12 & 1.05   \\
        & Liability         & 0.84 & 7.61 & 12.43 & -4.82  \\
        \addlinespace
        \multicolumn{6}{l}{\textbf{Trained on AI-AI}} \\
        & No Institution    & 0.38 & -0.79 & 10.00 & -10.79  \\
        & Verifiability     & 0.39 & 5.13 & 4.28 & 0.85  \\
        & Liability         & 0.83 & 13.74 & 6.12 & 7.62 \\
        \addlinespace
        \multicolumn{6}{l}{\textbf{Trained on Human-Human}} \\
        & No Institution    & 0.38 & -0.29 & 9.33 & -9.62   \\
        & Verifiability     & 0.43 & 4.82 & 5.30 & -0.48   \\
        & Liability         & 0.83 & 7.61 & 12.30 & -4.69  \\
        \bottomrule
    \end{tabular}
    \vspace{0.1cm}
    \caption*{\small \textit{Note:} Group-level summary statistics. \textit{Relative Efficiency} is calculated as the ratio of actual group income and maximum potential group income. All values are derived from group averages across 1000 simulations. \textit{Consumer Surplus} denotes the average cumulative income of all consumers in a group, \textit{Expert Surplus} denotes the average cumulative income of all experts in a group.}
    \label{tab:efficiency_hai}
\end{table}
This is in line with prior research about the importance of social preferences for improving market outcomes, as well as the human tendency to trust human agents more than AI agents. When consumers are protected from overcharging, LLMs can perform on par with human experts, but only with access to data from human-human interactions. Otherwise, they set prices that deter consumers, possibly because they do not (automatically) obfuscate their subsequent treatment intentions. Finally, a liability institution erases any difference between either the nature of the expert, the training regiment of the LLM, or respective prices. This is in line with theoretical considerations and past research showing liability to be very efficient in sustaining prosperous credence goods markets. Furthermore, it suggests that humans can, to some extent, be partially substituted without any welfare losses, if effective, targeted regulations (i) exist and (ii) are enforceable.

\noindent
\textbf{Market Efficiency and Surplus Distribution.} Table \ref{tab:efficiency_hai} shows summary statistics for the three training conditions and the three institutional frameworks.

Because we utilize a one-shot survey with a self-interested expert LLM that exhibits no variance in treatment and price charging behavior, we draw the consumer's problem type randomly across 1000 simulations. This minimizes the effect of problem assignment on market outcomes. First, note that in general, all three indicators follow a similar pattern as in the human market above. Market efficiency is strongly contingent on a liability rule, which also maximizes consumer and expert surplus. Between training conditions, the only significant effect on relative market efficiency occurs in the verifiability framework, where, in accordance with higher consumer approach rates, the human-trained LLM expert leads to significantly higher average group incomes than the untrained ($t = 5.63, \, p < 0.001$) and the AI-trained LLM ($t = 5.48, \, p < 0.001$). Furthermore, there are several distributive effects. Consumer surplus is larger in the \textbf{No Training} condition without any institutional rules ($t = 15.66, \, p < 0.001$; $t = 15.69, \, p < 0.001$), slightly lower in the verifiability condition when the LLM is trained on human data ($t = 5.19, \, p < 0.001$; $t = 5.28, \, p < 0.001$), and a lot larger for the AI-trained LLM under \textit{Liability} ($t = 41.72, \, p < 0.001$; $t = 41.22, \, p < 0.001$). These results follow basic pricing and consumer approach patterns. Without any institutions, the untrained LLM provides by far the lowest prices that offer consumers an expected value of 0. Under \textit{Verifiability}, consumers are more likely to approach the human-trained LLM because their prices signal equal markups. However, because the LLM is prone to undertreat consumers despite those markups, average surplus decreases. Finally, under liability, prices are by far the lowest in \textbf{AI Trained}, translating into strong consumer gains as experts are always forced to solve the problem. Note that, once again, the effect of the AI's training source on consumer surplus fundamentally depends on the credence goods market's institutional framework.

Now, we compare the hybrid expert market with the human expert market on our three indicators. To ensure comparability, we use the same rates of big and small problems as documented in the human market (Section 5.2) when calculating and comparing outcomes, instead of the simulated numbers. Table \ref{tab:comp_hh_hai} shows OLS regressions of treatment variables on relative efficiency, consumer surplus and expert surplus with the human market as a baseline.
\begin{table}[t]
\centering
\footnotesize
\caption{Comparison Human-Human and Human-AI Market}
\label{tab:comp_hh_hai}
\begin{tabular}{l@{\hspace{0.5em}}c@{\hspace{0.5em}}c@{\hspace{0.5em}}c@{\hspace{0.5em}}c@{\hspace{0.5em}}c@{\hspace{0.5em}}c@{\hspace{0.5em}}c@{\hspace{0.5em}}c@{\hspace{0.5em}}c}
\hline \hline
& \multicolumn{3}{c}{Relative Efficiency} & \multicolumn{3}{c}{Consumer Surplus} & \multicolumn{3}{c}{Expert Surplus} \\
& \multicolumn{1}{c}{(1)} & \multicolumn{1}{c}{(2)} & \multicolumn{1}{c}{(3)} & \multicolumn{1}{c}{(4)} & \multicolumn{1}{c}{(5)} & \multicolumn{1}{c}{(6)} & \multicolumn{1}{c}{(7)} & \multicolumn{1}{c}{(8)} & \multicolumn{1}{c}{(9)} \\
& No Inst. & Verif. & Liab. & No Inst. & Verif. & Liab. & No Inst. & Verif. & Liab. \\
\hline
\multicolumn{10}{c}{\textit{Baseline Category: Human-Human Market}} \\
[0.5em]
AI Training & -0.212*** & -0.264*** & 0.014 & -8.917*** & -4.523*** & \textcolor{blue}{2.853***} & \textcolor{blue}{6.023***} & \textcolor{blue}{0.648**} & 0.377 \\
& (0.026) & (0.028) & (0.033) & (0.638) & (0.632) & (0.328) & (0.427) & (0.234) & (0.697) \\
Human Training & -0.216*** & -0.234*** & 0.020 & -8.349*** & -4.851*** & -3.284*** & \textcolor{blue}{5.334***} & \textcolor{blue}{1.668***} & \textcolor{blue}{6.545***} \\
& (0.026) & (0.028) & (0.033) & (0.636) & (0.630) & (0.326) & (0.426) & (0.233) & (0.694) \\
No Training & -0.227*** & -0.268*** & 0.025 & -4.507*** & -4.561*** & -3.280*** & \textcolor{blue}{1.191**} & \textcolor{blue}{0.506*} & \textcolor{blue}{6.615***} \\
& (0.026) & (0.028) & (0.033) & (0.637) & (0.631) & (0.327) & (0.426) & (0.233) & (0.695) \\
\hline
Observations & 907 & 907 & 907 & 907 & 907 & 907 & 907 & 907 & 907 \\
\hline \hline
\multicolumn{10}{p{0.95\textwidth}}{\footnotesize Notes: The table reports OLS regression results with treatment dummies. Standard errors in parentheses. * p<0.05, ** p<0.01, *** p<0.001. Blue coefficients indicate significant positive effects.} \\
\end{tabular}
\end{table}
Compared to the human market, the hybrid market is less efficient across all training and institution conditions, except \textit{Liability}. Note that, for \textbf{Human Training} and \textit{Verifiability}, that cannot be explained by the number of successful interactions, but is instead caused by expert fraud. Columns (4) -- (6) document significant and substantial decreases in consumer earnings across all specifications, except when the LLM expert is both liable and trained on simulated data. Expert surplus increases concurrently, albeit not enough to compensate the consumer losses. These results suggest that, in the absence of an enforceable liability institution, introducing generative AI into markets for expert services can reduce overall efficiency by substantially decreasing consumer earnings. Moreover, even when experts are liable, LLMs tend to re-distribute profits from consumers to experts. This can have an equalizing effect (Figure \ref{fig:surplus_gap} in the Appendix), but reduces total income significantly. In sum, generative AI hurts overall welfare without an adequate liability rule. It re-distributes economic surplus from consumers to experts, who benefit across the board, potentially leading to distorting incentives. 

\section{Expert Delegation and Human-AI-Human Interactions}
We have shown that LLM experts reduce overall welfare in the absence of a liability rule through substantial consumer losses and a reduction in successful interactions. However, despite net losses, experts, on average, generate significantly higher surpluses, potentially incentivizing LLM adoption. Furthermore, simulations confirm that changes in the LLM's objective function can categorically shift AI treatment choices. Therefore, this section analyzes whether human experts are willing to delegate pricing and treatment choices on credence goods markets to a LLM, how the ability to influence the AI's objective function changes delegation behavior, and which objectives experts prefer. In addition, we gather experimental data on Human-AI-Human expert markets in which human consumers interact with human experts who either (1) did or did not choose to delegate their pricing and treatment choices to an LLM or (2) did or did not choose their LLM's social preference objective function, which are -- depending on treatment -- transparent or obfuscated to consumers.

\subsection{Experimental Design}
The expert procedure follows the one described above. Participants complete six within-subject condition (No Institution vs. Verifiability vs. Liability) x (Fixed Objective vs. Chosen Objective) of the credence goods one-shot experiment in randomized order. The only difference is that now, experts have the option to delegate their price-setting, diagnosis and treatment choices to an LLM-agent who can earn money on their behalf. In \textbf{Fixed Objective}, experts learn that the LLM has been prompted with the objective to ``maximize Player A's payoff''. In addition, they learn the LLM is Claude 3.5 Sonnet by Anthropic, that it has learned from instructions that are very similar to the ones provided to human experts, and that all Player B's observe whether an expert chose to delegate before making their approach decision. Like consumers, experts receive an information box with more detailed information about Claude 3.5 Sonnet. In \textbf{Chosen Objective}, we inform experts that, after deciding whether to rely on the LLM agent, they will be allowed to choose one of four LLM objective prompts to ``guide them in their decision making''. This is the only difference.\footnote{Note that experts do not learn that the LLM's objective function may be transparent to consumers. The main reason is that we are primarily concerned with experts delegation patterns, how they are shaped by the ability to affect the AI-agent's objective, and which social preferences experts exhibit. Inducing consumer transparency adds a strategic layer to the expert's decision, which is an interesting point for follow-up research.}

\noindent
\textbf{Consumer Procedure.} For consumers, the procedure mirrors the one from the Human-AI Interactions above, with a few distinct differences. First, consumers learn that each expert has the choice to delegate their price-setting, diagnosis and treatment to the LLM agent. Second, when observing expert prices and making an approach decision, consumers also learn whether an expert delegated to the AI system, or relied on their own choices. Here, we introduce a 2 (Transparency vs. No Transparency) x 2 (Fixed Objective vs. Chosen Objective) between-subject design. In \textbf{No Transparency}, consumers observe prices and delegation choices, but learn nothing about the LLM's objective function. In \textbf{Transparency}, consumers learn that the LLM's objective is to maximize Player A's payoff (\textbf{Fixed Objective}), or which of the four possible objective prompts the expert chose (\textbf{Chosen Objective}). Like before, we implement the three institutional variations as within-subject conditions. All observations are gathered from \textit{Connect}, restricted to US-based participants with an approval rating of at least 90\% (female = 46\%). We pre-registered to collect data until 150 independent observations per between-subject condition, with a base-payment of \$2. In total, we gathered 610 observations.

\subsection{Results -- Human-AI-Human Interaction}
\begin{figure}[h]
    \centering
    \caption{Left: Expert LLM Delegation Rates in \textbf{Fixed Objective} and \textbf{Chosen Objective}. Right: Share of Expert Social Preference Types Following their Chosen LLM Objective.}
    \includegraphics[width=0.49\textwidth]{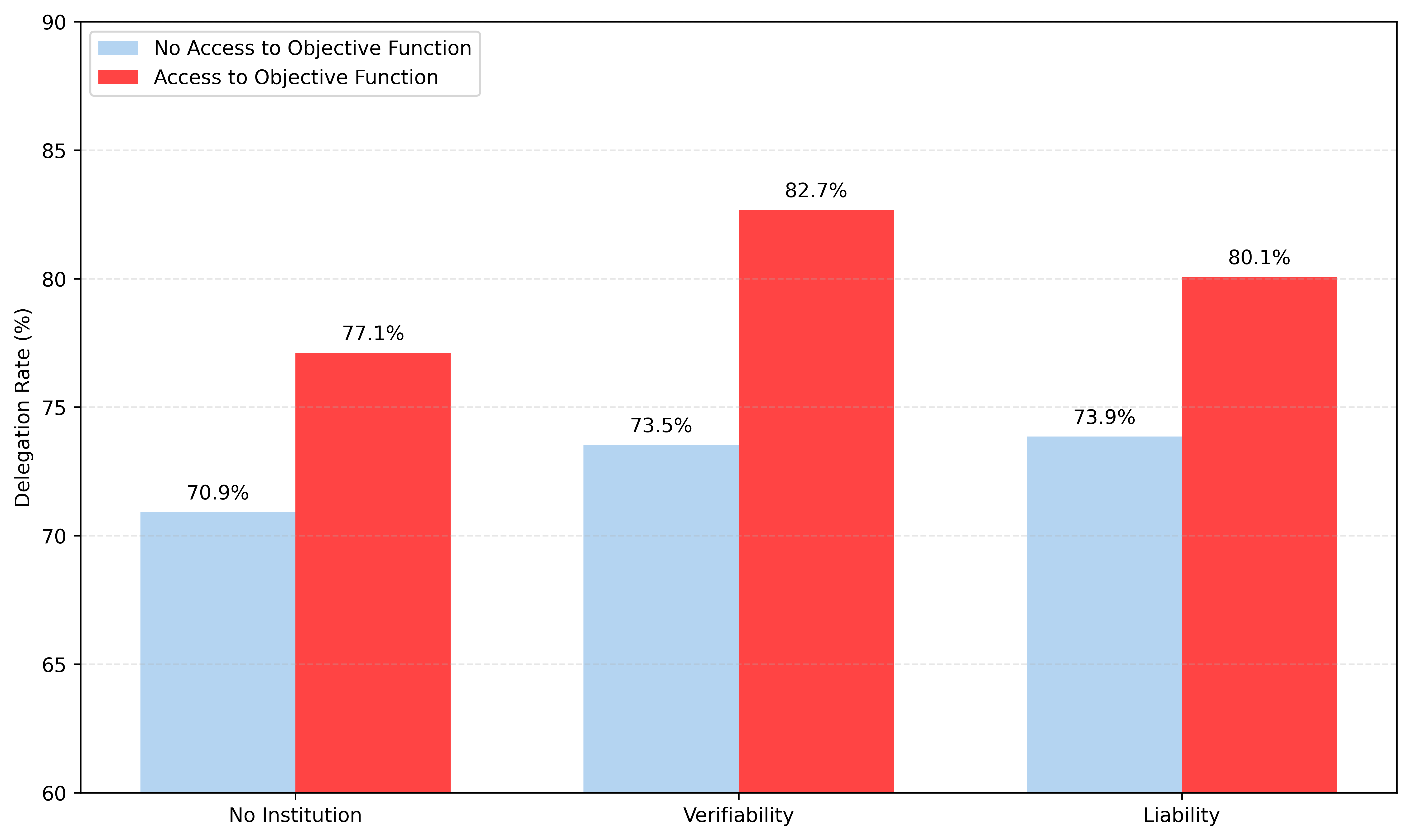}
    \includegraphics[width=0.49\textwidth]{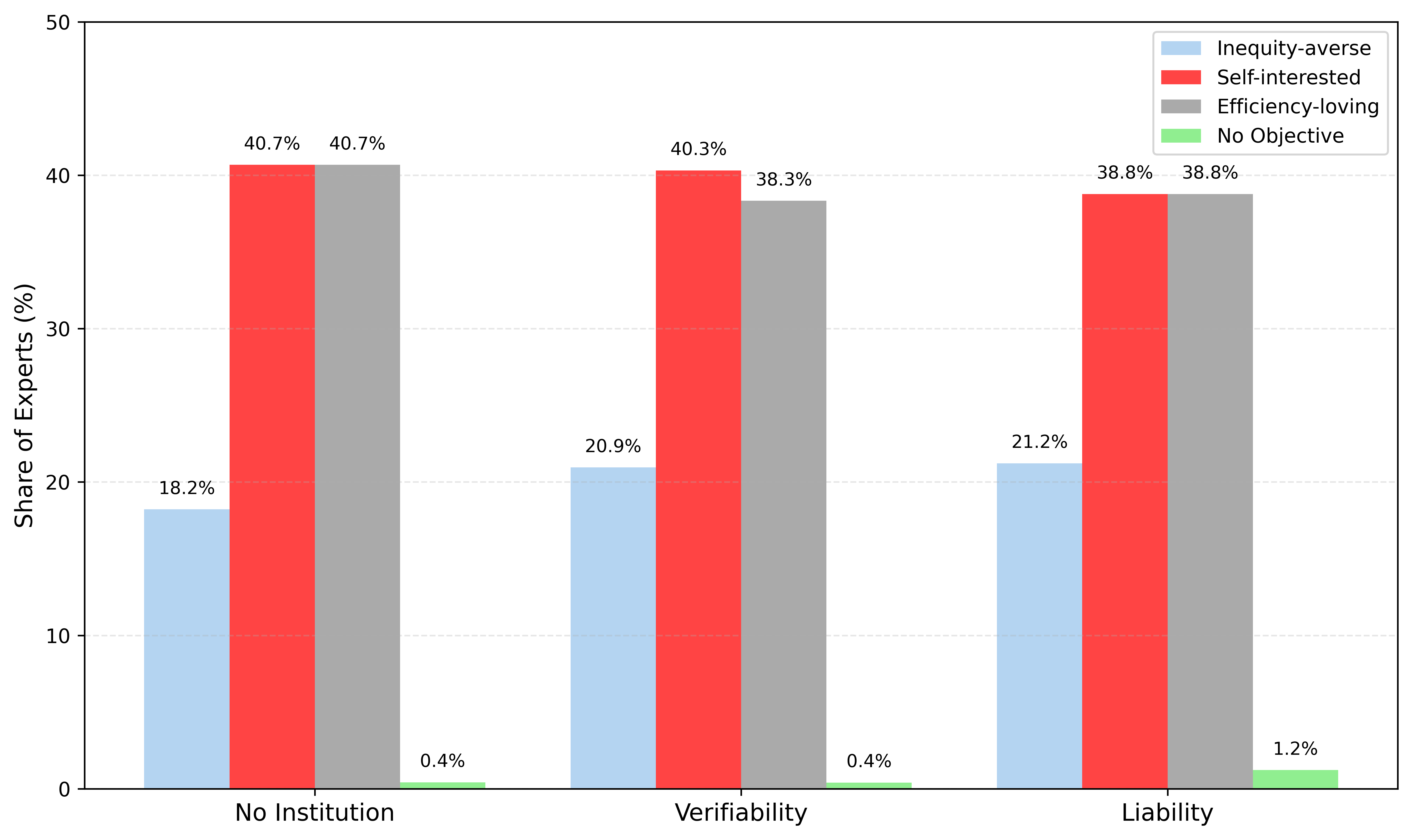}
    \label{fig:expert_delegation}
\end{figure}

Figure \ref{fig:expert_delegation} shows the share of experts delegating their decisions to the LLM in \textbf{Fixed Objective} and \textbf{Chosen Objective}. 

Reliance is high across all conditions, with a large majority of experts using the LLM independent of the institutional environment. Moreover, endowing participants with the option to subsequently choose between four objective prompts significantly increases reliance (see Table \ref{tab:reg_exp_del}). After delegation, the most popular objectives are \textit{self-interested} and \textit{efficiency-loving}, while 20\% opt for the inequity-averse model. A small majority therefore codifies preferences that also consider the consumer's payoff, although overall, economic considerations are much more important than fairness concerns. Furthermore, there are no difference in objective preferences between institutions, confirming that experts do not appear to predict how social preferences and institutions interact (see the AI-AI Interactions in Section 5.1). Regarding human expert fraud, there are no differences between the objective treatments, while overall shares are similar to those documented on the Human-Human market (Figure \ref{fig:expert_objective_fraud} in the Appendix), suggesting no selection effects. Overall and in line with the re-distribution of economic surpluses documented in Human-AI markets above, most human experts delegate their decisions to the LLM. 

\begin{figure}[t]
    \centering
    \caption{Left: Aggregated Consumer Approach Decisions across Institutional Frameworks. Right: Consumer Approach Decisions when all Experts Delegate.}
    \includegraphics[width=0.49\textwidth]{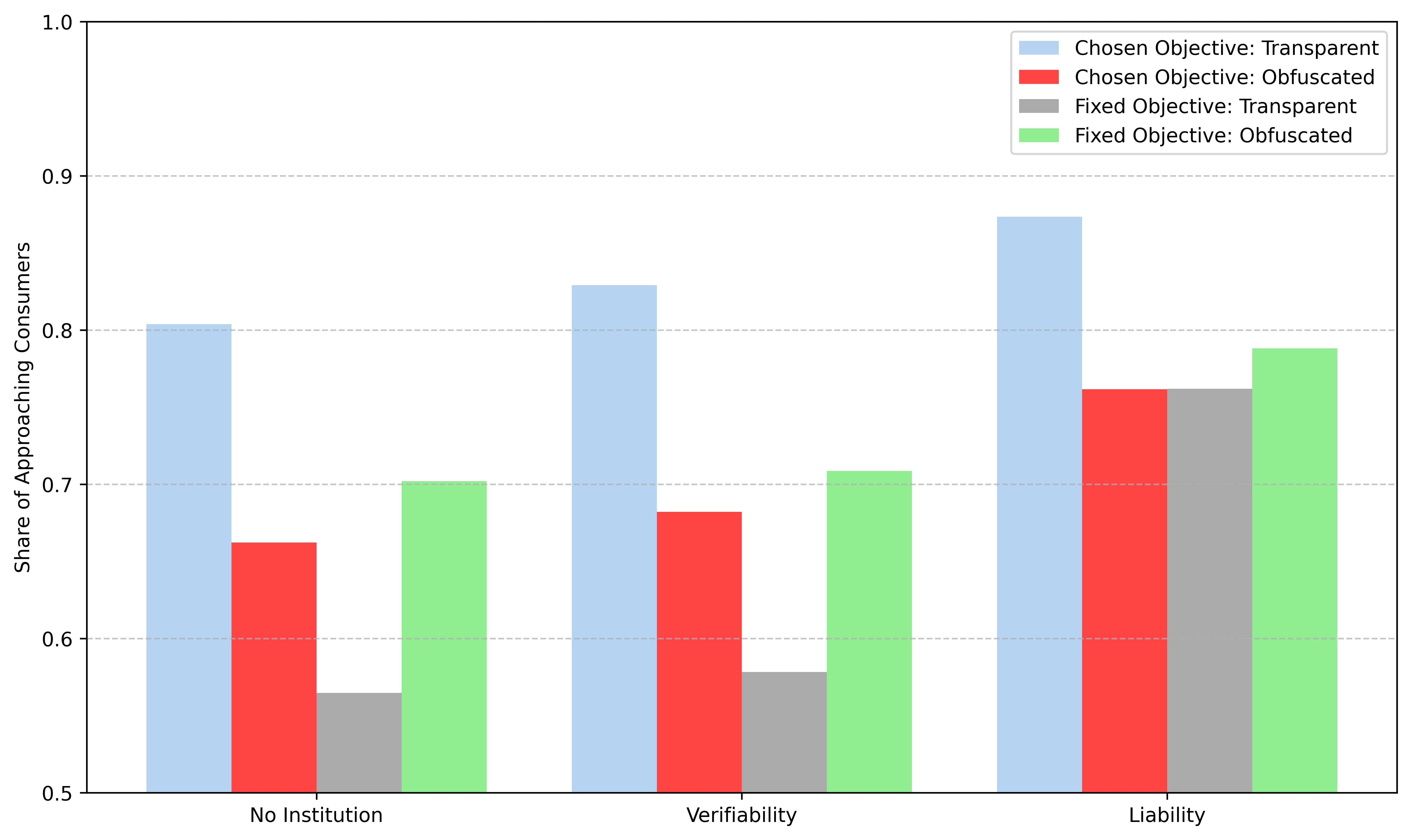}
    \includegraphics[width=0.49\textwidth]{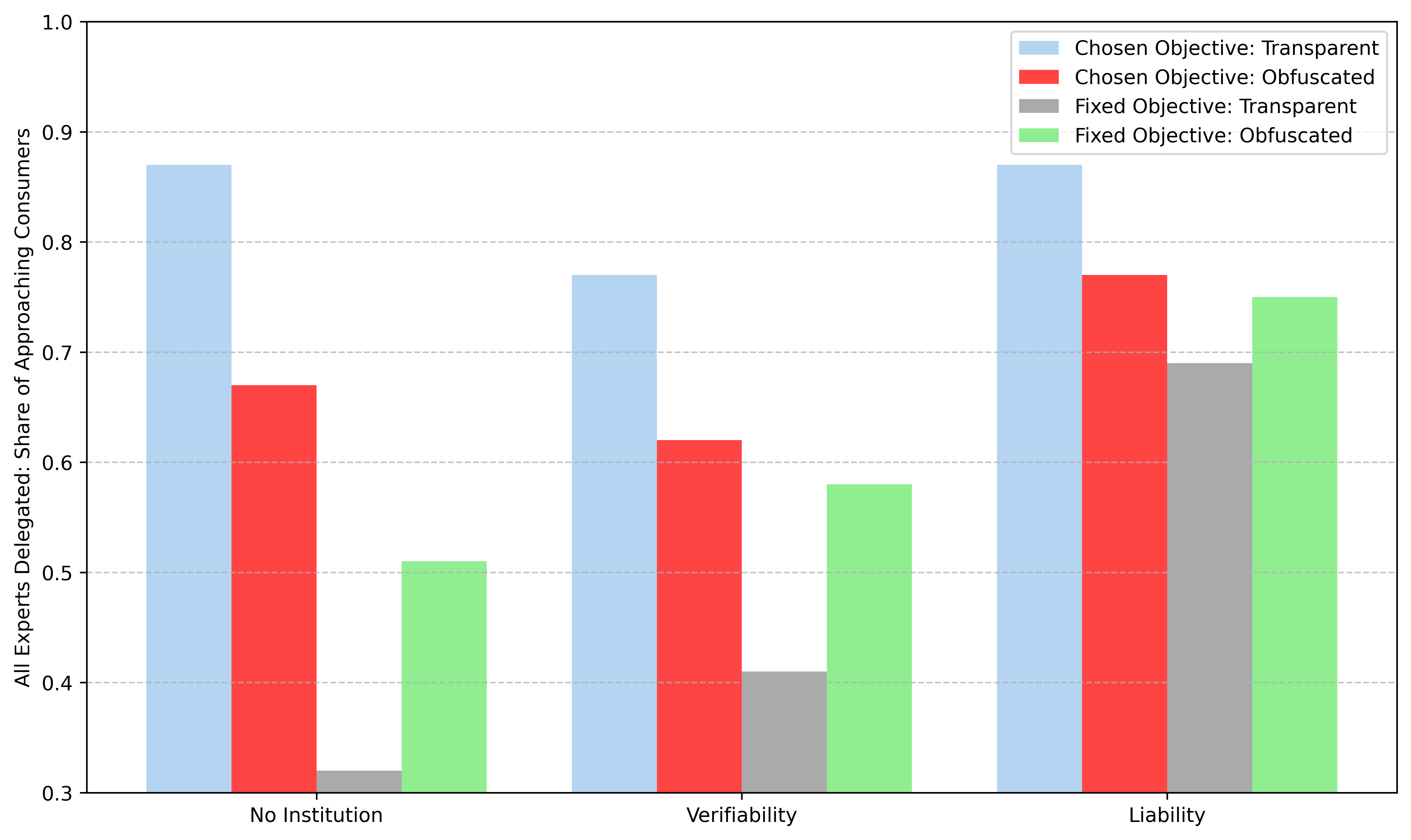}
    \label{fig:consumer_approach_haih}
\end{figure}

\noindent
\textbf{Consumer Behavior.} Introducing human experts and LLM objective functions into Human-AI markets changes consumer reactions.\footnote{LLM treatment decisions are equivalent to the AI-AI interactions. LLM prices in \textit{No Institution} are 3/5, 4/8, 4/8 for self-interested/efficiency-loving/inequity-averse; in \textit{Verifiability} they are 4/7, 4/8, 4/8 and in \textit{Liability} 4/8, 4/8, 4/8 respectively.} The left panel of Figure \ref{fig:consumer_approach_haih} shows average consumer approach rates by treatment and institution. Consumer behavior significantly differs between treatments in \textit{No Institution} ($\, \tilde{\chi}^2 = 20.78, p < 0.001$), \textit{Verifiability} ($\, \tilde{\chi}^2 = 23.28, p < 0.001$) and \textit{Liability} ($\, \tilde{\chi}^2 = 8.08, p = 0.044$). This is driven by two main effects. One, when experts can choose between different LLM objective functions that codify social preferences, consumers are significantly more likely to approach an expert if the objective function is transparent (\textit{No Institution:} 0.80 vs. 0.66, $\, \tilde{\chi}^2 = 7.94, p = 0.005$; \textit{Verifiability:} 0.83 vs. 0.68, $\, \tilde{\chi}^2 = 9.07, p = 0.003$; \textit{Liability:} 0.87 vs. 0.76, $\, \tilde{\chi}^2 = 6.51, p = 0.011$). In contrast, if the LLM by default maximizes the expert's payoff, transparency significantly hurts consumer attraction in the absence of liability (\textit{No Institution:} 0.56 vs. 0.70, $\, \tilde{\chi}^2 = 7.94, p = 0.005$; \textit{Verifiability:} 0.58 vs. 0.71, $\, \tilde{\chi}^2 = 9.07, p = 0.003$; \textit{Liability:} 0.76 vs. 0.79, $\, \tilde{\chi}^2 = 0.29, p = 0.588$). If we focus on consumer choices when all four experts choose to delegate their price-setting and treatment choices to the LLM (right panel of Figure \ref{fig:consumer_approach_haih}), the patterns are qualitatively analogous.\footnote{\textbf{Chosen Objective:}(\textit{No Institution:} 0.87 vs. 0.67, $\, \tilde{\chi}^2 = 7.19, p = 0.007$; \textit{Verifiability:} 0.77 vs. 0.62, $\, \tilde{\chi}^2 = 3.59, p = 0.058$; \textit{Liability:} 0.87 vs. 0.77, $\, \tilde{\chi}^2 = 1.87, p = 0.171$). \textbf{Fixed Objective:} (\textit{No Institution:} 0.32 vs. 0.51, $\, \tilde{\chi}^2 = 3.04, p = 0.081$; \textit{Verifiability:} 0.41 vs. 0.58, $\, \tilde{\chi}^2 = 2.75, p = 0.097$; \textit{Liability:} 0.69 vs. 0.75, $\, \tilde{\chi}^2 = 0.35, p = 0.551$).} Moreover, the rates in \textbf{Fixed Objective} with transparent objective functions closely align with the approach rates documented in the Human-AI Interactions. Obfuscating the self-interested motive appears to increase successful human-expert interactions without liability rules, which decreases average consumer surplus, because they are less likely to avoid fraudulent LLM treatment choices. Finally, note that allowing experts to choose an objective function does not change consumer approach rates compared to self-interested LLM experts when the objective function remains obfuscated. As shown below, this can cause substantially higher expert profits under the fixed self-interested regime, at the expense of consumer surplus, highlighting the problem of adverse expert incentives without transparency rules.

Which objective function do consumers favor? We can compare approach rates between \textbf{Transparency} and \textbf{No Transparency} for \textbf{Chosen Objective} to gauge which objective function ``over-performs'' relative to a blind scenario in which the only differences are the prices (Figure \ref{fig:consumer_obj_Haih} in the Appendix). The most prominent pattern is that in both \textit{No Institution} and \textit{Verifiability}, the most successful objective function switches from ``self-interested'' under obfuscation to ``efficiency-loving'' with transparency. That is likely because the self-interested LLM sets lower prices than the other artificial experts. In particular, the share of consumers choosing a self-interested LLM expert falls from 53\%, 49\% and 25\% to 11\%, 14\% and 9\% across the three respective institutional environments. In contrast, the ``efficiency-loving'' objective function increases consumer attraction from 20\%/15\%/27\% to 58\%/44\%/41\%. Similarly, transparency increases the attractiveness of the ``inequity-averse'' objective function, whereas a \textit{lower} share of consumers opt for the human non-delegating expert. This is in line with our predictions above, whereby transparent (and consequential) social preferences attract more consumers by reducing the information asymmetry. Transparency substantially increases the success of LLM experts that consider the payoffs of both players, whereas self-interested LLMs in particular, but also human experts, attract less consumers. 

What about the \textbf{Fixed Objective} conditions? Because all LLMs exhibit a fixed ``self-interested'' objective function, we are primarily interested in consumer preferences between human expert and human experts who delegated to the LLM. In line with intuition, we find the largest effect for \textit{No Institution}, where 20\% of consumer subjects approach a human expert under obfuscation, which increases to 53\% once the LLM's self-interested objective function is revealed ($\, \tilde{\chi}^2 = 21.31, p < 0.001$). Notably, while there is a similar trend for the other institutional conditions, differences are smaller and not significant (\textit{Verifiability}: 0.47 vs. 0.56; \textit{Liability}: 0.40 vs. 0.49). Again, the self-interested LLM expert in \textit{No Institution} sets lower prices than most human experts, which appears to attract more consumers when the objective function is obfuscated, despite still being theoretically less appealing than the outside option. Overall, consumers strongly condition their approach behavior on the LLM's objective function, whereas the ``type'' of expert (human vs. LLM) has much less explanatory power. In particular, experts attract the most consumers if they (i) can choose the LLM's objective function, (2) opt for ``efficiency-loving'' or ``inequity-averse'' preferences, and (3) reveal these preferences to consumers. On the other hand, disclosing objectives makes self-interested expert fraud mostly nonviable through competition. In line with theory and prior results, the largest gains in terms of successful expert-consumer interactions happen when full expert liability is absent or non-enforceable. 

\begin{table}[t]
    \small	
    \centering
    \caption{\textit{No Institution.} Average Market Efficiency, Consumer Surplus and Expert Surplus}
    \begin{tabular}{lcccc}
        \toprule
        Treatment & Relative Efficiency & Consumer Surplus & Expert Surplus & $\Delta$ \\
        \midrule
        Chosen Objective: Transparent & 0.74 & 2.9 & 1.5 & 1.4 \\
        Chosen Objective: Obfuscated & 0.52 & 1.7 & 1.4 & 0.3 \\
        Fixed Objective: Transparent & 0.52 & 2.1 & 1.1 & 1 \\
        Fixed Objective: Obfuscated & 0.51 & 1.2 & 1.9 & -0.7 \\        
        \bottomrule
    \end{tabular}
    \vspace{0.1cm}
    \caption*{\small \textit{Note:} Group-level summary statistics. \textit{Relative Efficiency} is calculated as the ratio of actual group income and maximum potential group income. All values are derived from group averages. \textit{Consumer Surplus} denotes the average income per consumer in a group, \textit{Expert Surplus} denotes the average income per expert in a group.}
    \label{tab:efficiency_noinst}
\end{table}

\noindent
\textbf{Market Efficiency and Surplus Distribution.} Table \ref{tab:efficiency_noinst} shows efficiency and surplus estimates for \textit{No Institution}. See Table \ref{tab:efficiency_veri} and Table \ref{tab:efficiency_lia} in the Appendix for \textit{Verifiability} and \textit{Liability}. We derive several insights, conditional in the institutional environment. One, in \textit{No Institution}, the combination \textbf{Chosen Objective} and \textbf{Transparency} delivers significantly and substantially higher efficiency than all other variants (Chosen\_Obfuscated: $t = 2.98, \, p = 0.004$; Fixed\_Transparent: $t = 2.88,\, p = 0.005$; Fixed\_Obfuscated: $t = 2.86, \, p = 0.005$), mostly due to increases in consumer surplus. The other three combinations are equivalent. Throughout, in the absence of institutional rules, experts earn the most with self-interested LLMs and obfuscation (Chosen\_Transparent: $t = 2.07, \, p = 0.04$; Chosen\_Obfuscated: $t = 2.24, \, p = 0.03$; Fixed\_Transparent: $t = 3.72, \, p < 0.001$). However, if transparency is mandatory, allowing experts to modify the LLM's objective function by including more pro-social preferences increases expert surplus through higher consumer participation ($t = 2.04, \, p = 0.04$). These experts simultaneously out-compete (a) human experts and (b) delegating experts who codify self-interest.    

Second, under \textit{Verifiability}, the same general efficiency pattern emerges. Revealing experts' objective choices leads to significantly higher efficiency than the three other treatments (Chosen\_Obfuscated: $t = 2.97, \, p = 0.004$; Fixed\_Transparent: $t = 3.38,\, p = 0.001$; Fixed\_Obfuscated: $t = 2.09, \, p = 0.039$), which do not differ from one another. Expert surplus is highest in \textbf{Chosen\_Transparent}, rather than \textbf{Fixed\_Obfuscated} (Chosen\_Obfuscated: $t = 3.75, \, p < 0.001$; Fixed Transparent: $t = 4.72,\, p < 0.001$; Fixed\_Obfuscated: $t = 4.09, \, p < 0.001$). This is largely driven by the LLM's effectiveness under transparent ``inequity-averse'' and ``efficiency-loving'' preferences. Hence, expert incentives \textit{without transparency rules} flip compared to \textit{No Institution}.

Third and in line with prior results, for \textit{Liability}, there are no significant differences between treatments. As one would suspect from above -- but not from the standard model with social preferences -- efficiency levels across the board are higher than those in the other institutional environments. However, qualitatively, allowing experts to commit to social preferences via LLM delegation comes close. In fact, the efficiency jump from either fixed or non-disclosed LLM social preferences to disclosed LLM social preferences without liability amounts to roughly 20 percentage points (or ca. 40\%), compared to a 35 percentage point gain through liability. Our results highlight important interactions between expert agency over an LLM's objective function, market institutions, and transparency rules. Most notably, without effective institutions and transparency rules, experts are incentivized to choose self-interested objective functions and obfuscate them. If objective functions are transparent, experts who use LLMs and codify self-interest are out-competed by other experts that signal more pro-social objectives. In the absence of liability, this pattern drives large efficiency gains, which either accrue to consumers (\textit{No Institution}) or both sides (\textit{Verifiability}).

\section{Discussion}
This paper derives empirical insights for AI-AI, Human-Human, Human-AI and Human-AI-Human expert markets. Four experiments consistently illustrate the importance of institutions for market efficiency and the distribution of economic gains. Crucially, we tease-out context-specific effects that depend on the interaction between (1) an AI's training data, or (2) transparency rules, and the installed institutional rule. Furthermore, we show that Human-Human markets generally outperform Human-AI and AI-AI markets in the absence of liability, while experts consistently benefit from deploying LLMs. In line with these results, a large majority of experts chooses to rely on LLMs. Giving experts the opportunity to codify their LLM's social preferences using prompts further increases delegation rates. Consumers are more likely to engage with LLM experts if they can observe that human experts codified efficiency-loving or inequity-averse social preferences. This drives substantial efficiency gains and reduces fraud. Consequently, Human-AI-Human markets can outperform the other three alternatives. However, in the absence of transparency, experts may still be incentivized to delegate their choices to self-interested LLMs, strongly reducing efficiency and hurting consumer surplus.

\noindent
\textbf{Behavioral Implications.} In choice environments that are characterized by information asymmetries between agents, social preferences can play a crucial role in fostering successful interactions and cooperation. For expert markets, two of the most important factors are consumer trust, which is often a necessary condition for successful interactions in the absence of perfect liability rules, and experts' other-regarding preferences towards consumers. To achieve sustained market activity, expert fraud must be relatively rare. As exemplified by our Human-Human one-shot market and prior experimental results, both consumers and experts often achieve higher rates of market activity through consumer trust and costly expert honesty. By comparing Human-Human and Human-AI markets, we have shown that consumers are c.p. less likely to approach an LLM expert that plays on behalf of a human expert. This pattern holds in all institutional environments where trust is consequential, but fades under liability, providing strong evidence for a loss in consumer trust. Hence, on its own, the dissemination of LLM experts has the potential to hurt the efficiency of expert markets through a change in social preferences, thereby magnifying the negative effect of prevailing information asymmetries. The market becomes less \textit{behavioral}. On the other side, the LLM tested in this study opts to defraud consumers by default and when instructed to maximize expert payoff. This causes higher rates of expert fraud compared to Human-Human markets, as the LLM does not mimic average expert preferences. The impact of experts' other-regarding preferences is crowded-out by monetary self-interest, increasing short-run profits and aligning outcomes with the standard model. 

On the other hand, allowing human experts to make (1) a delegation choice and (2) an LLM preference choice \textit{increases} the positive effect of social preferences on expert market outcomes under certain conditions. While a large group of experts opt for a self-interested LLM agent, even more experts consider consumers in their objective function, most prominently through an efficiently-loving LLM agent that maximizes the total payoff of both agents. In the presence of LLMs, most experts exhibit other-regarding preferences. If these preferences are public information, other-regarding experts who rely on LLMs out-compete both self-interested LLM experts and human experts, leading to (1) higher market activity and (2) higher consumer surplus. By revealing a codified objective function, experts can reduce the information asymmetry while simultaneously signaling other-regarding preferences, which increases consumer trust and therefore consumer participation. Here, transparency is a necessary condition, as without it, consumers do not exhibit higher levels of trust. Without transparency, experts who do not choose to maximize their own self-interest are disadvantaged. This means that LLMs can substantially alleviate the credence goods problem by increasing the effectiveness of social preferences, but \textit{only} if experts can credibly reveal their codified preferences or the LLM's codified objective function. Note that, in the absence of regulations, consumers will probably mostly judge such disclosure as cheap talk, as there are no incentives for self-interested experts to honestly reveal their objectives.

Finally, we provide behavioral evidence that experts are more likely to rely on an LLM if they have agency over the LLM's objective function. This result relates to the literature on algorithm aversion, which has documented a positive relationship between expert agency and algorithm utilization \citep{burton2020systematic}. Specifically, \cite{dietvorst2018overcoming} show that people are more likely to use prediction algorithms if they can modify their output -- even if the adjustments are very small. In this paper, we extend this result to (1) LLMs with (2) a high baseline utilization rate and (3) human agency over the AI agent's input -- the LLM's objective function --, rather than their output, in (4) strategic interactions.

\noindent
\textbf{AI-AI Interactions.} Expert competition can theoretically clear credence goods markets under simple game-theoretic assumptions of self-interest. Therefore, it would be plausible that generative AI can overcome the information asymmetry problem via incentive-compatible pricing. In contrast, results show that when both consumers and experts are represented by LLMs, there are no successful market interactions without expert liability, irrespective of the LLM's objective function. This is mainly due to three reasons. One, LLMs set prices that are too high, and generally do not address consumer concerns about potential expert fraud. Second, LLM consumers are risk-averse, and only approach LLM experts if it is the dominant choice. Third, LLM consumers do not extend any trust to LLM experts. Therefore, only liability solves the credence goods problem on one-shot AI markets. This is also important because we find that LLM experts adjust treatment and price-charging behavior on explicit objective prompts -- which are by default not transparent to consumers, but have the potential to substantially increase market efficiency. Self-interested LLMs and those without any explicit objective consistently opt to dishonestly defraud consumers, vindicating the latter's trust and risk preferences. Introducing inequity-averse or efficiency-loving preferences generally shifts the LLM towards honesty, which allows for long-term mutually beneficial economic interactions. This, however, does not hold for (1) an inequity-averse expert whose choices are verifiable or (2) an efficiency-loving expert who is liable. Hence, our results highlight the complex interplay of social preferences and institutions, even when no human decision-maker is involved. They also suggest that LLMs may assist consumers to evade expert fraud. LLM consumers are generally able to identify that incentives are misaligned, and consequently make very risk-averse choices. This disproportionately benefits liability regulations, as they become a necessary condition for the market to exist. Moreover, because LLM consumers always approach liable experts, social preferences become a ``mere'' tool to redistribute surpluses. This not only makes the deployment of LLMs safer, but also simpler. The market always produces maximum welfare, while competition disciplines price-setting. Social planers can still make distributive changes by inducing social preferences, but should do so with caution, and a comprehensive understanding of the institutional framework.

\noindent
\textbf{Human-Human Interactions.} Our results from one-shot Human-Human interactions are mostly in line with the existing experimental literature on credence goods markets. 
Compared to AI-AI interactions, market efficiency is much higher in \textit{No Institution} and \textit{Verifiability}, but lower in \textit{Liability}. Price competition leads to, on average, low prices, which is why consumers always generate between 50\% and 70\% more surplus than experts. Still, the overwhelming majority of experts is hesitant to offer theoretically predicted sub-cost prices, and the few who do are not necessarily more successful. Compared to AI-AI markets, human markets have a large comparative advantage when the institutional setting is not robust to expert fraud. As long as experts are not disincentivized to, e.g., mis-treat or overcharge consumers, there is no AI consumer demand for expert services, and AI agents always defraud consumers in the absence of explicit social preference codification. In contrast, human participants are more efficient through both demand-side and supply-side effects. The majority of consumers lends experts enough trust to approach them, which is re-paid through relatively low rates of expert fraud. As a consequence, both consumer and expert surplus is higher. These results relate to prior work showing how social preferences within human-human interactions can increase market efficiency above what would be expected by standard game-theoretical considerations operating under the assumption of self-interest \citep{fehr2003nature}. Hence, if liability institutions do not exist, are too costly, or cannot be enforced, delegating human decisions on expert markets to LLM agents severely hurts overall welfare. Here, human social preferences function as a substitute institution, imperfectly compensating for the lack of institutional support. However, if liability \textit{can} be reliably installed and enforced, full strategic delegation -- under the assumption of an exogenous correct diagnostic signal -- increases welfare on both sides. That is because even under liability, roughly 20\% of human consumers refrain from approaching a human expert. LLMs ``solve'' this cooperation problem by following predictable, consistent game-theoretic choice patterns.

\noindent
\textbf{Human-AI and Human-AI-Human Interactions.} Human consumers that seeks advice or services from LLMs are already ubiquitous, and this relationship will only intensify with the advance of LLM-based products, apps, services and advisory chat bots. Our results about the effects of substituting human experts with LLM agents that decide on behalf of human experts are ambivalent, showing heterogeneous effects across institutional and training regimes. 

In general, similar to humans, LLMs struggle to set prices that are theoretically desirable. Pricing depends on training data, and using simulated data from AI-AI interactions leads to unusual price patterns. Specifically, the LLM fails to maximize expert income by setting very low prices in \textit{Liability}, reveals their subsequent dishonest strategy in \textit{Verifiability}, and opts for a price pair in \textit{No Institution} that only 3.2\% of human experts choose. In contrast, using data from human expert markets aligns the LLM with generally successful pricing behavior from human experts, while simultaneously avoiding competition prices that drive profits down to 0. It uses the price pair with the highest success rate that implies some expert profits under honesty in \textit{No Institution}, commits to equal markups in \textit{Verifiability}, and chooses the most prominent price pair in \textit{Liability}. This has some positive consequences for both expert and overall welfare, albeit only in \textit{Verifiability}. Two straightforward implications are (1) using simulated interaction data can be a harmful training regime in complex strategic economic settings that involve information asymmetries and (2) learning from past human data allows the LLM to learn and apply human heuristics, with potentially positive implications.

Looking at consumer behavior, we again document nuanced effects. Without any institution, approach shares are significantly lower across the three training conditions. This is despite the extremely low prices in \textbf{No Training}, suggesting that consumer choices are sensitive to expert incentives. For both the off-the-shelve LLM and the AI-trained LLM, consumers continue to leave the market at higher rates in \textit{Verifiability}. Hence, a dissemination of LLMs across expert markets can lead to welfare losses. The LLM in \textbf{Human Trained} is competitive with human experts. Intuitively, setting prices that erase the economic incentive for dishonesty alleviates human distrust into LLMs, equalizing the rate of successful interactions. As long as incentives are present, however, consumers avoid the LLM expert. Note that, in one-shot situations, this may be both beneficial or harmful to consumer welfare, as consumers generally have no insight into the expert's specific objective function. If the LLM has no explicit objective to care about the consumer, these patterns can prevent consumer exploitation. Yet, if it does, a lack of trust hurts both experts and consumers. This supports the assertion that specifically in expert markets, where the existent information asymmetry between consumers and experts makes the utility of transactions highly uncertain, social preferences -- and expectations about social preferences -- can be crucial in sustaining market activity, but are partially crowded-out by the introduction of generative AI. 

Overall, changes in price-setting, treatment and approach behavior cause two dominant effects when going from Human-Human to Human-AI markets. One, \textit{overall welfare decreases in the absence of a liability rule}. This is primarily driven by lower rates of market interaction and higher rates of expert under-treatment. Two, irrespective of overall market efficiency, \textit{expert earnings generally increase at the expense of consumer income}. This has three main implications. First, experts are, on average, incentivized to rely on self-interested AI, despite rampant consumer defrauding, questionable price-setting, and more untreated consumers. Supply-side AI agent markets allow experts to escape price competition and exploit consumers via proxy. Second, codifying and enforcing expert liability may even be more important with the dissemination of generative AI. Third, when pro-social preferences are important for market efficiency, LLMs can have unintended negative consequences specifically by crowding-out the ``human element''. 

Fortunately, our Human-AI-Human experiment reveals one potential solution to all these problems: allowing expert to induce social preferences into the LLM's objective function \textit{and} revealing these preferences to consumers causes a substantial jump in efficiency, with particularly large gains for consumers. There are two main channels. One, other-regarding LLM experts out-compete selfish LLM experts (and humans with obfuscated preferences). This leads to high market participation, and a low probability of expert fraud. Two, experts themselves benefit from using other-regarding preferences, alleviating adverse incentives. However, without transparency, social preferences have no positive effect on market efficiency or consumer surplus. In contrast, self-interested LLM experts earn higher payoffs for their human experts, re-introducing the aforementioned incentive problems. 

These results highlight the potential role of policy-makers and regulations for achieving efficient expert-consumer interactions. In a competitive expert setting, transparency about an AI model's objectives functions like the proverbial ``free lunch''. One, both other-regarding and self-interested experts cannot afford to codify preferences that signal potential fraud, as consumers strongly condition their choices on the observed objective function. Second, signaling non-selfish preferences allows experts to attract more consumers, increasing their own payoff. Overall, transparency reduces information asymmetries, as experts commit a priori to a treatment strategy. Without transparency rules, however, social preferences \textit{do not} out-compete self-interested LLMs. Furthermore, in the absence of credible commitments (like regulations), all experts are incentivized to signal other-regarding preferences, irrespective of their true choice. This removes the information value of transparency and, hence, reverts the market back to a standard credence goods market. Therefore, we argue, transparency and explicit expert adjustments of the LLM's objective or reward function should be seriously discussed. This directly feeds into current discussion about the \href{https://artificialintelligenceact.eu/}{EU AI Act's} optimal scope, but could also be relevant for national regulatory bodies like the FDA in the context of medical AI. Note that the consequences of codifying such ``social concerns'' into an LLM's objective function can be dependent on the institutional framework, and should therefore not be centrally mandated across different industries or domains. This holds especially in light of likely future consumer segmentation, where AI experts end up serving consumers with standardized needs, while humans focus on complex cases, or those consumers who value social interaction. Hence, it is likely that AI systems predominantly disseminate for standardized services, which (1) are easier to regulate and (2) allow for AI-targeted approaches.

\clearpage
\bibliography{\bib}

\clearpage
\section*{Appendix}

\begin{table}[h!]
    \centering
    \caption{Expert Price Setting Across Treatments}
    \begin{tabular}{l@{\hspace{1em}}ccc@{\hspace{1.5em}}ccc@{\hspace{1.5em}}ccc}
        \toprule
        & \multicolumn{3}{c}{No Institution} & \multicolumn{3}{c}{Verifiability} & \multicolumn{3}{c}{Liability} \\
        \cmidrule(lr){2-4} \cmidrule(lr){5-7} \cmidrule(lr){8-10}
        Treatment & $\ubar{p}$ &  $\bar{p}$ & N &  $\ubar{p}$ &  $\bar{p}$ & N &  $\ubar{p}$ &  $\bar{p}$ & N \\
        \midrule
        No Objective      & 4.0 & 5.0 & 160 & 4.0 & 7.0 & 160 & 5.0 & 7.0 & 160 \\
        Maximize Payoff   & 5.0 & 8.0 & 160 & 5.0 & 5.0 & 160 & 4.0 & 7.0 & 160 \\
        Inequity Averse   & 6.0 & 8.0 & 160 & 5.0 & 5.0 & 160 & 4.0 & 8.0 & 160 \\
        Efficiency Loving & 4.0 & 8.0 & 160 & 4.0 & 8.0 & 160 & 4.0 & 8.0 & 160 \\
        \bottomrule
    \end{tabular}
    \vspace{0.1cm}
    \caption*{\small \textit{Note:} Model: claude-3-5-sonnet-20241022. Prices by LLM experts depending on the prompted objective function across different market institutions in the AI-AI-Interaction framework.}
    \label{tab:expert_prices}
\end{table}

\begin{table}[h]
\centering
\caption{Most Frequent Expert Price Pairs by Treatment}
\label{tab:price_pairs_rankings}
\begin{tabular}{ccccccc}
\hline
\multicolumn{2}{c}{No Institution} & \multicolumn{2}{c}{Verifiability} & \multicolumn{2}{c}{Liability} \\
Price Pair & \% & Price Pair & \% & Price Pair & \% \\
\hline
(4, 8) & 17.30 & (2, 6) & 15.98 & (4, 8) & 20.00 \\
(3, 7) & 14.75 & (4, 8) & 15.82 & (3, 7) & 12.79 \\
(2, 6) & 14.59 & (3, 7) & 12.87 & (2, 6) & 12.46 \\
(5, 8) & 4.67 & (5, 9) & 6.31 & (5, 8) & 5.16 \\
(3, 6) & 3.93 & (5, 10) & 4.02 & (5, 9) & 4.92 \\
(5, 10) & 3.77 & (3, 6) & 3.85 & (6, 8) & 3.44 \\
(4, 7) & 3.20 & (5, 8) & 3.28 & (5, 10) & 3.36 \\
(5, 7) & 3.20 & (4, 7) & 3.11 & (6, 10) & 2.62 \\
(6, 8) & 2.79 & (4, 9) & 2.70 & (4, 10) & 2.38 \\
\hline
\end{tabular}
\end{table}

\clearpage

\clearpage

\begin{figure}[h]
    \centering
    \caption{$\Delta$ of Consumer and Expert Surplus in the Human Market and the Hybrid Markets.}
    \includegraphics[width=0.9\textwidth]{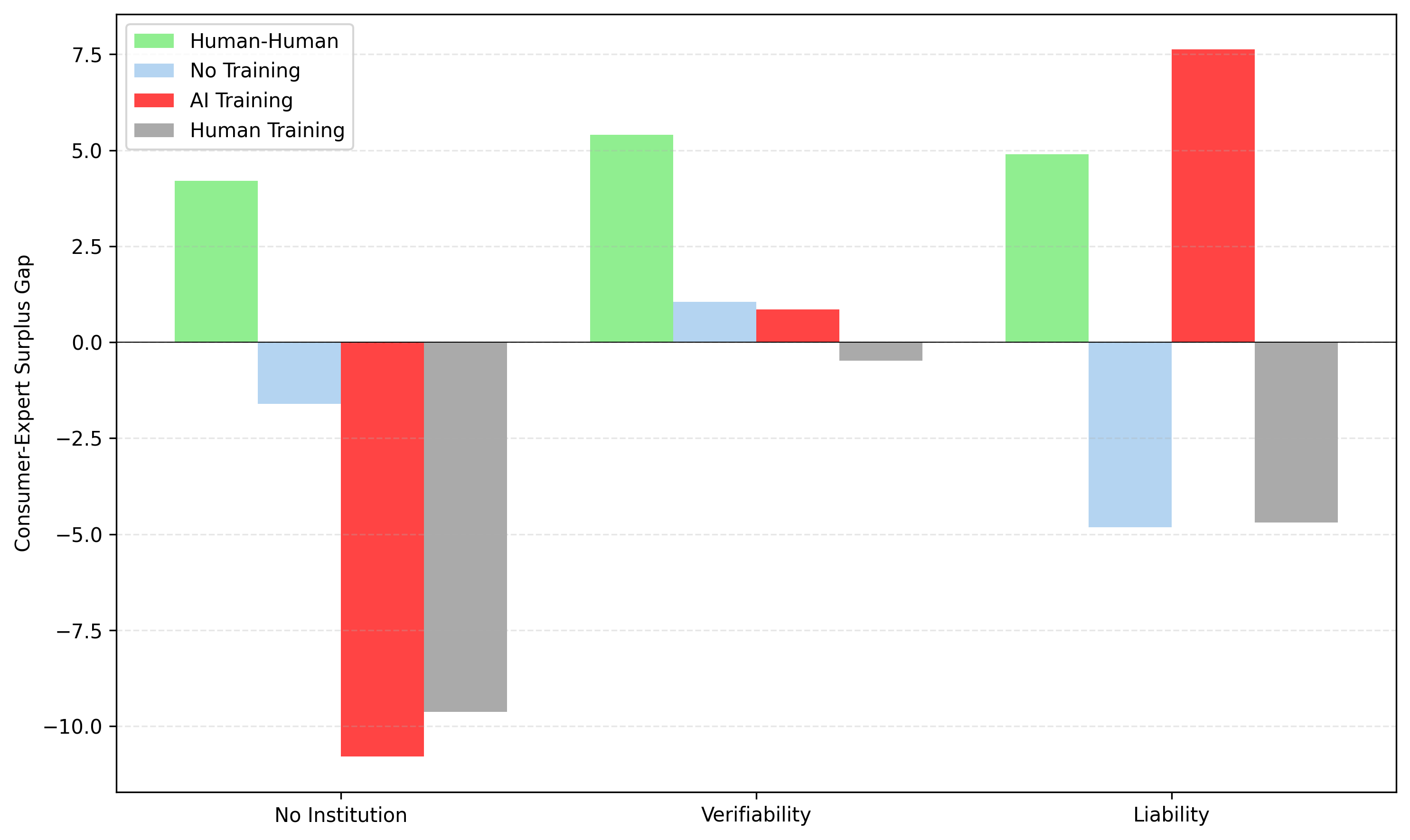}
    \label{fig:surplus_gap}
\end{figure}

\begin{figure}[h]
    \centering
    \caption{Left: Expert Fraud Shares \textbf{No Objective}. Right: Expert Fraud Shares in \textbf{Objective}.}
    \includegraphics[width=0.49\textwidth]{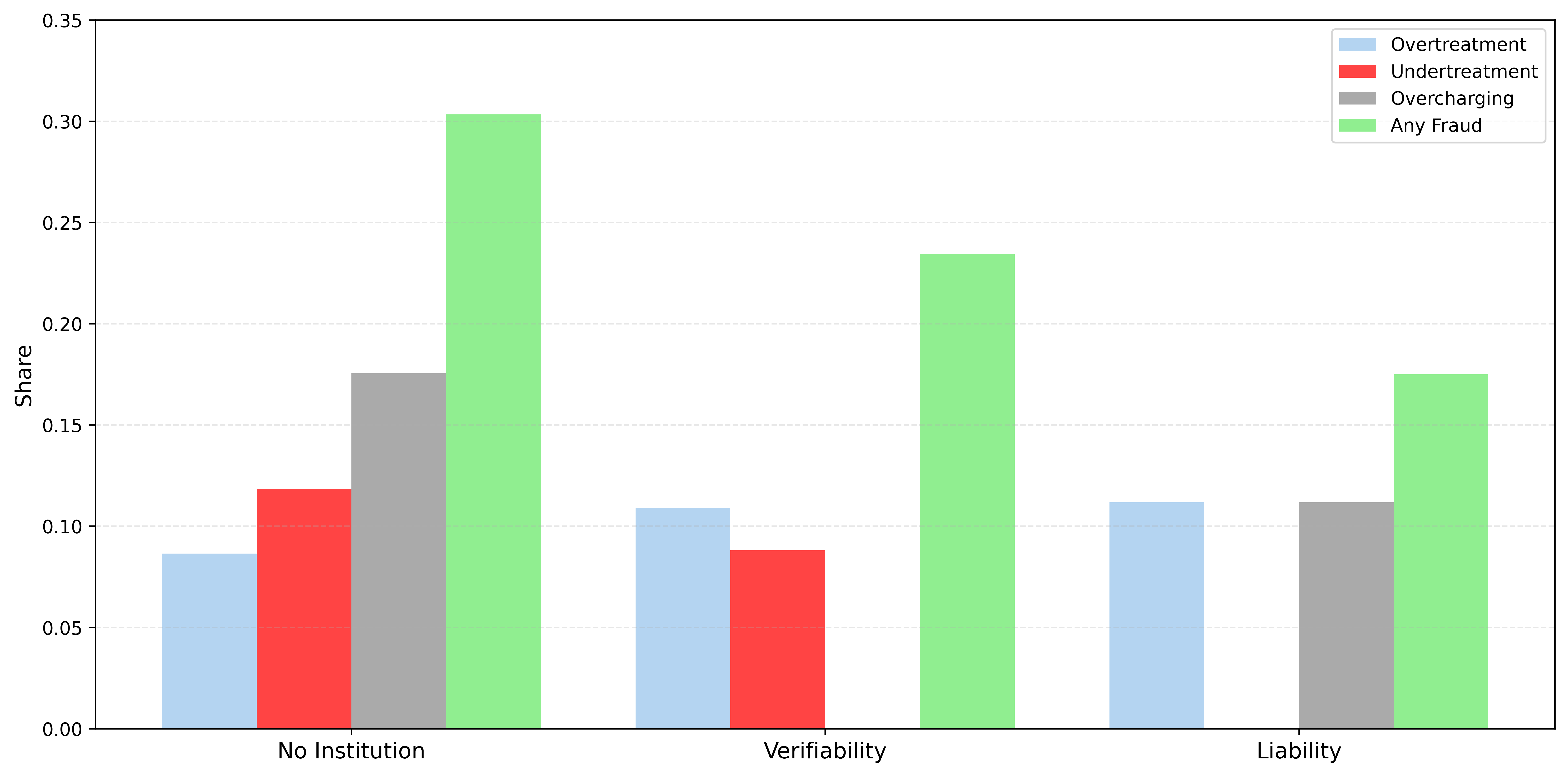}
    \includegraphics[width=0.49\textwidth]{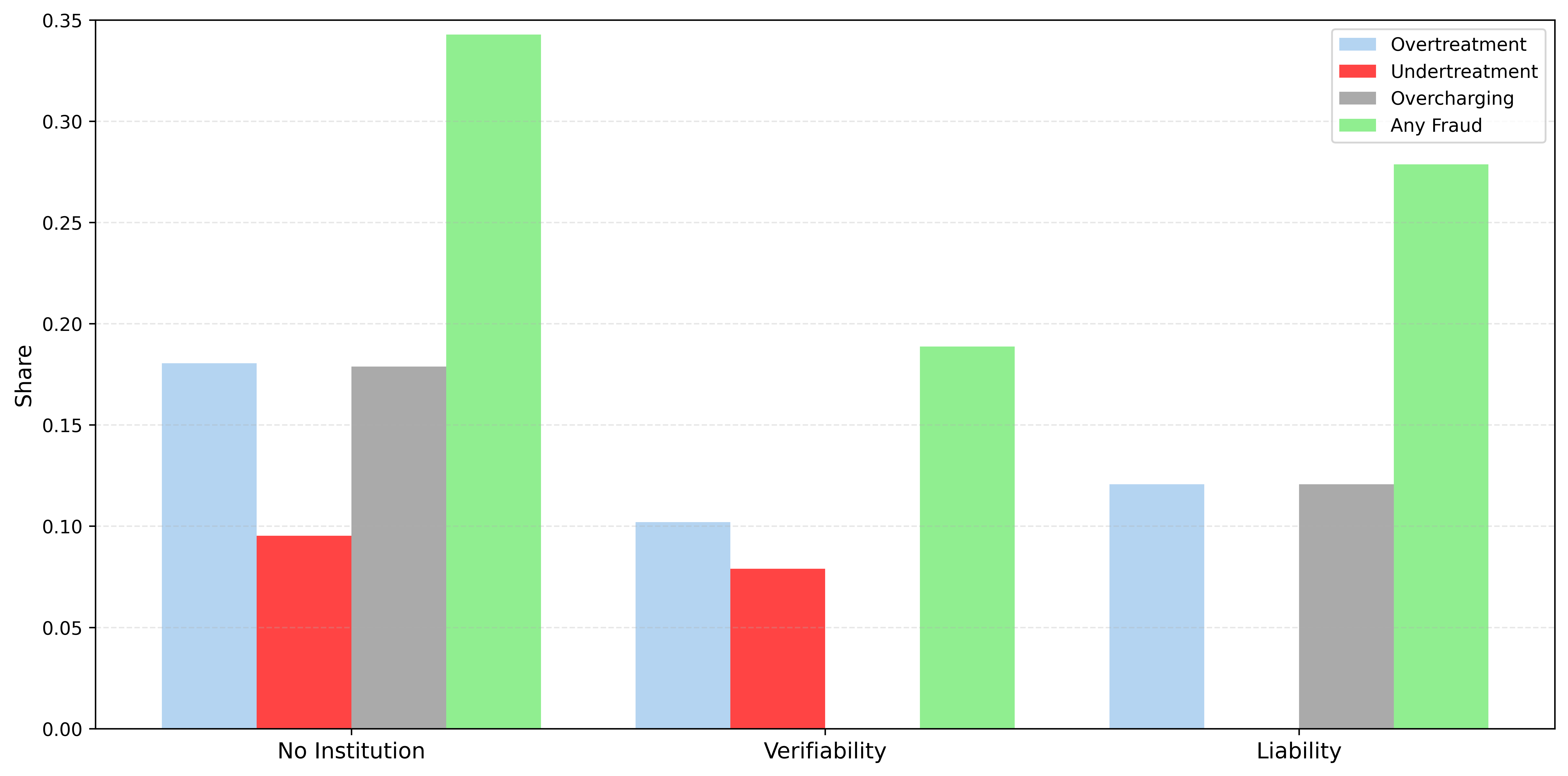}    
    \label{fig:expert_objective_fraud}
\end{figure}

\begin{table}[h]
\centering
\small
\caption{Expert Delegation Behavior}
\label{tab:reg_exp_del}
\begin{tabular}{lcccccc}
\hline \hline
& \multicolumn{6}{c}{Repeated Measures Logit} \\
& (1) & (2) & (3) & (4) & (5) & (6) \\
& No Institution & No Institution & No Verifiability & Verifiability & Liability & Liability \\
\hline
Chosen Objective & 0.06* & 0.06* & 0.09*** & 0.09*** & 0.06* & 0.06* \\
& (0.03) & (0.03) & (0.02) & (0.02) & (0.02)  & (0.02) \\
Risk & & 0.02 & & 0.08 & & 0.008 \\
& & (0.01) &  & (0.01) & & (0.01)\\
Age & & 0.005$^{**}$ &  & 0.004$^{*}$ & & 0.005$^{***}$ \\
& & (0.00) &  & (0.00) &  & (0.00)\\
Female & & 0.031 &  & 0.095$^{*}$ &   &  0.06\\
& & (0.039) &  & (0.04) &  & (0.04) \\
\hline
Observations & 612 & 612 & 612 & 612 & 612 & 612 \\
\hline \hline
\multicolumn{7}{l}{\footnotesize Notes: Table shows population-averaged effects from GEE with unstructured correlation.}\\
\multicolumn{7}{l}{\footnotesize Dependent variable: binary variable capturing whether an expert delegated to the LLM.} \\
\multicolumn{7}{l}{\footnotesize Standard errors in parentheses. $^{*}$ p$<$0.05, $^{**}$ p$<$0.01, $^{***}$ p$<$0.001} \\
\end{tabular}
\end{table}

\begin{figure}[h]
    \centering
    \caption{Approached Objective Function in the \textbf{Chosen Objective} Treatments.}
    \includegraphics[width=\textwidth]{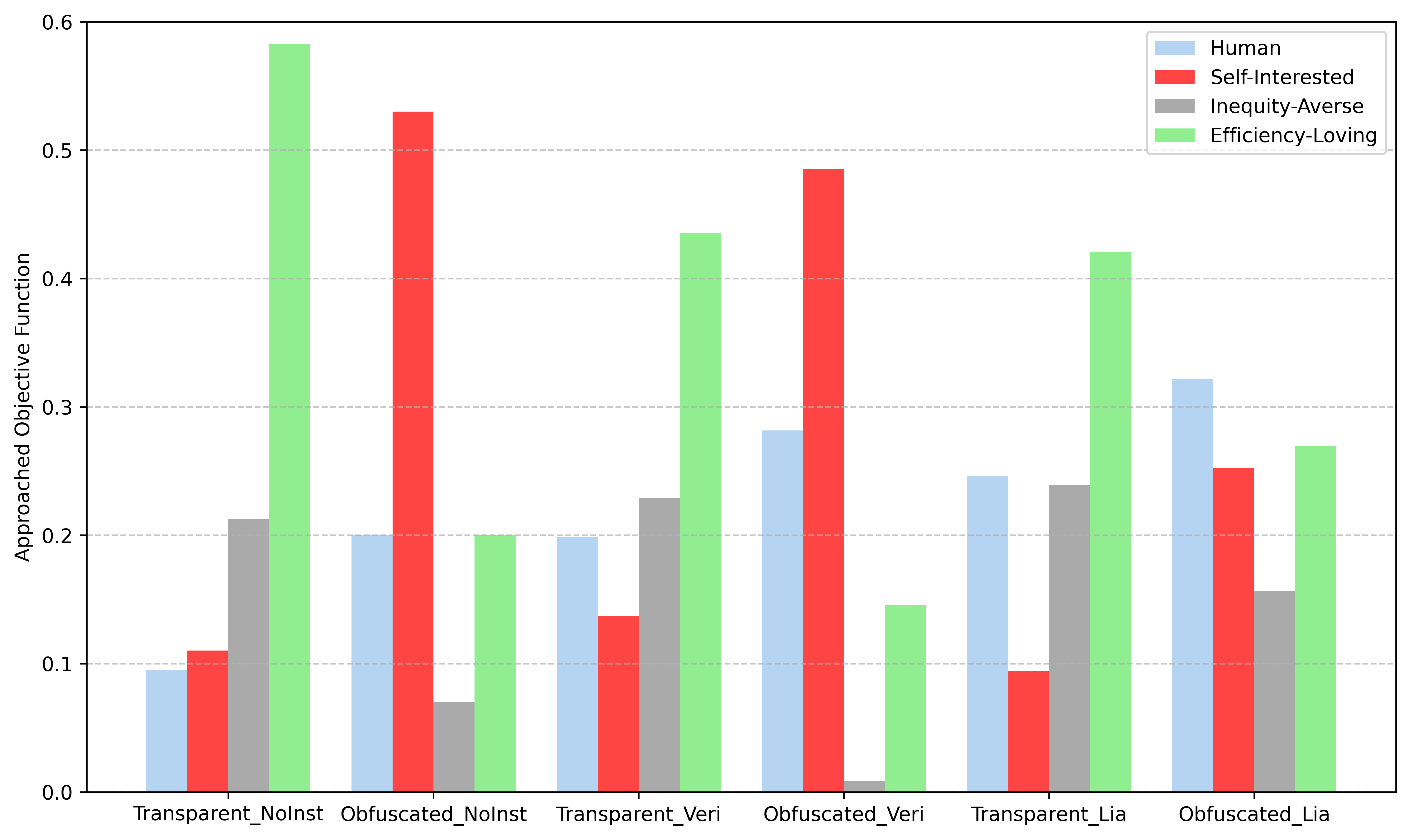}
    \label{fig:consumer_obj_Haih}
\end{figure}

\begin{table}[h]
    \small	
    \centering
    \caption{\textit{Verifiability.} Average Market Efficiency, Consumer Surplus and Expert Surplus}
    \begin{tabular}{lcccc}
        \toprule
        Treatment & Relative Efficiency & Consumer Surplus & Expert Surplus & $\Delta$ \\
        \midrule
        Chosen Objective: Transparent & 0.72 & 2.8 & 1.5 & 1.3 \\
        Chosen Objective: Obfuscated & 0.52 & 2.1 & 1.0 & 1.1 \\
        Fixed Objective: Transparent & 0.47 & 2.1 & 0.7 & 1.4 \\
        Fixed Objective: Obfuscated & 0.56 & 2.5 & 0.9 & 1.6 \\        
        \bottomrule
    \end{tabular}
    \vspace{0.1cm}
    \caption*{\small \textit{Note:} Group-level summary statistics. \textit{Relative Efficiency} is calculated as the ratio of actual group income and maximum potential group income. All values are derived from group averages. \textit{Consumer Surplus} denotes the average income per consumer in a group, \textit{Expert Surplus} denotes the average income per expert in a group.}
    \label{tab:efficiency_veri}
\end{table}

\begin{table}[h]
    \small	
    \centering
    \caption{\textit{Liability.} Average Market Efficiency, Consumer Surplus and Expert Surplus}
    \begin{tabular}{lcccc}
        \toprule
        Treatment & Relative Efficiency & Consumer Surplus & Expert Surplus & $\Delta$ \\
        \midrule
        Chosen Objective: Transparent & 0.94 & 3.4 & 2.3 & 1.1 \\
        Chosen Objective: Obfuscated & 0.83 & 3.0 & 1.9 & 1.1 \\
        Fixed Objective: Transparent & 0.85 & 3.1 & 2.0 & 1.1 \\
        Fixed Objective: Obfuscated & 0.91 & 3.3 & 2.2 & 1.1 \\        
        \bottomrule
    \end{tabular}
    \vspace{0.1cm}
    \caption*{\small \textit{Note:} Group-level summary statistics. \textit{Relative Efficiency} is calculated as the ratio of actual group income and maximum potential group income. All values are derived from group averages. \textit{Consumer Surplus} denotes the average income per consumer in a group, \textit{Expert Surplus} denotes the average income per expert in a group.}
    \label{tab:efficiency_lia}
\end{table}

\end{document}